\documentclass[a4paper,12pt]{article} 
\pdfoutput=1
\usepackage{jcappub}
\usepackage[latin1]{inputenc}
\usepackage{dsfont} 
\usepackage[english]{babel}
\usepackage{bm}
\usepackage{dsfont} 

\usepackage[normalem]{ulem}
\newcommand{\eref}[1]{Eq.~(\ref{#1})}
\newcommand{\fref}[1]{Figure~\ref{#1}}
\newcommand{\sref}[1]{Section~\ref{#1}}

\title{
Effect of extended $\nu$ production region on collective oscillations in supernovae
}

\author[a,b]{Rasmus S. L. Hansen}
\author[a,c]{and Alexei Yu. Smirnov}
\affiliation[a]{Max-Planck-Institut f\"ur Kernphysik,\\ Saupfercheckweg 1, 69117 Heidelberg, Germany}
\affiliation[b]{Department of Physics and Astronomy, Aarhus University, \\ 
Ny Munkegade 120, 8000 Aarhus C, Denmark}
\affiliation[c]{Korea Institute for Advanced Study, 85 Hoegiro, Seoul 02455, Republic of Korea}
\emailAdd{rshansen@phys.au.dk}
\emailAdd{smirnov@mpi-hd.mpg.de}

\abstract{
In supernovae neutrinos are emitted from a region with a width $r_{\rm eff}$
of a few kilometers (rather than from a surface of infinitesimal width). 
We study the effect of integration (averaging) over such an extended emission region
on collective oscillations. 
The averaging leads to additional suppression of the correlation
(off-diagonal element of the density matrix)  by a factor $ \sim 1/r_{\rm eff} V_e  \sim 10^{-10}$, 
where $V_e$ is the matter potential. 
This factor  enters the initial condition for further collective oscillations  
and, consequently,  leads to a delay of the strong flavour transitions. 
We justify and quantify this picture using a simple example  of 
collective effects in two intersecting fluxes. 
We have derived the evolution equation for the density matrix elements integrated over 
the emission region and solved it both numerically and analytically. 
For the analytic solution we have used linearized equations. 
We show that the delay of the development of the instability and the collective oscillations 
depends on the suppression factor due to the averaging (integration) logarithmically. 
If the instability develops inside the production region, the integration leads 
not only to a delay but also to a modification of the exponential grow.  
} 

\begin{document}

\maketitle

\section{Introduction}
\label{sec:intro}

In most studies of the neutrino flavour conversion and, 
in particular, effects of collective oscillations, 
the emission of neutrinos is considered 
from a neutrino sphere -- a surface of zero width,  
so that further neutrino evolution proceeds in the free streaming regime. 
The radius of the neutrino sphere $R_{\nu \rm sphere}$ 
is defined in terms of the optical depth $\tau (r)$ as~\cite{Keil:2002in}
\begin{equation}
  \label{eq:Rnusphere}
  \tau(R_{\nu \rm sphere}) = \frac{2}{3},
\end{equation}
where
\begin{equation}
  \label{eq:opticaldepth}
  \tau(r) \equiv  \int_r^\infty \Gamma(r')dr',  
\end{equation}
and $\Gamma$ is the collision rate.
Thus, $\tau$ is the probability  of collision from the emission point $r$ to 
$\infty$, and according to (\ref{eq:Rnusphere})  and  (\ref{eq:opticaldepth}),  
 $R_{\nu \rm sphere}$ is the radius from 
which the probability for neutrinos to escape without collisions is 1/3. 

This zero width approximation was justified
by the fact that in the production region, the matter density
is very large. Hence the in matter mixing angle is small 
and, consequently, any flavour evolution are strongly suppressed. 
Therefore, the flavour evolution was considered from some surface 
outside the production region~\cite{Pantaleone:1994ns}. 

In reality, the region of neutrino production $r_{\rm eff}$
is rather large  (we quantify this in \sref{sec:effwidth}).   
In \cite{Hansen:2018apu} it was shown that the zero width approximation may not be correct:
in spite of the smallness of mixing, the 
integration over the neutrino production region
leads to additional suppression of flavour transitions.
According to \cite{Hansen:2018apu}, strong transitions can
be interpreted as a parametric effect driven by
periodic modulations of the background potentials.
The bigger the amplitude of modulations, the faster the transition.
Therefore any averaging, which diminishes the
depth of modulations $d$, will suppress the transition.
The spatial scale of strong transitions $L_{\rm par}$ is inversely proportional to
the depth: $L_{\rm par} \propto 1/d$. Consequently,  the averaging leads to an increase of $L_{\rm par}$,
{\it i.e.}, delay  of the transition. 
Since the potential induced by neutrinos (which drives collective
oscillations) decreases quickly, a delay may mean  that  the transition
never occurs. Here the delay is linear in $V_e r_{\rm eff}$,  
where $V_e$ is the matter potential, 
since the instability develops linearly with distance.

The solvable example considered in \cite{Hansen:2018apu} has one important
shortcoming: $\nu \nu$ interactions
in the background neutrino flux are neglected. With $\nu \nu$ interactions being included, 
parametric effects can be more complicated and the instability develops much faster. 

In this paper, we will explore effects of
averaging over the neutrino production region 
in the presence of $\nu \nu$ interactions in the background. 
Our approach is the following: We study the integration effect using a simple and symmetric model in which collective effects show up. Namely, the model with two intersecting beams with a single energy and equal numbers of neutrinos and antineutrinos. Then we argue that the results obtained in this model are generic and to some extent should be reproduced in more realistic models. 
We show  that the integration (averaging)
leads to the suppression of the initial flavour transition by a factor $1/V_e r_{\rm eff}$.  
This suppression in turn leads to a delay (in distance) of the development of the instability and strong collective effects. 
The delay is given by the logarithm of the suppression factor:  $\log (1/V_e r_{\rm eff})$. 
If the instability starts in the production region, the averaging also modifies  
the exponential growth of the transition.  

The paper is organised as follows. In \sref{sec:effwidth} we  
make realistic estimations of the effective width of the production region $r_{\rm eff}$ and 
compute suppression of the conversion in the case of negligible 
collective  oscillations in the production region which is justified in the very early phases of evolution.  
In \sref{sec:simple} we consider the simplest system of two intersecting neutrino fluxes 
in which $\nu \nu$ interactions are included and study the effect of integration over the production region. 
In \sref{sec:linear} using the linearised evolution equations, we give an analytic 
description of the conversion. 
The analytic results are in very good agreement with the results of numerical computations. 
In \sref{sec:getreal}, using the analytic formulas, we extrapolate the results to parameters that are more realistic for the situation in supernovae. 
We conclude in \sref{sec:conclusions}. 
Some details of the computations are given in the appendices.

\section{Effective emission width and suppression of the initial correlation}
\label{sec:effwidth}

In this section we consider realistic conditions (density profiles) in supernovas,   
but neglect effects of $\nu \nu$ interactions, which is justified  in the  early 
phases of the flavour evolution (see \sref{sec:linear}).  
Indeed, at high densities close to the proto-neutron star, collective oscillations 
mainly manifest themselves as synchronised oscillations~\cite{Pastor:2001iu}. 
The exception is the very fast conversions~\cite{Sawyer:2005jk, Sawyer:2008zs, Sawyer:2015dsa, Chakraborty:2016lct, Dasgupta:2016dbv,Capozzi:2018clo,Shalgar:2019kzy} 
that might be relevant, e.g., when neutrinos are emitted asymmetrically 
from the supernova~\cite{Dasgupta:2016dbv} or in the late stages of the SN evolution~\cite{Shalgar:2019kzy}.
Furthermore, if the conditions for fast conversions are not satisfied, 
it is a good approximation to consider neutrinos with just one energy 
and ignore the neutrino background when estimating the effects of an extended neutrino sphere. 
This corresponds to the background neutrinos considered in \cite{Hansen:2018apu}. 

Notice that the single energy approximation has some similarity to the small amplitude synchronised oscillations\footnote{An important difference is that synchronised oscillations lead neutrinos and antineutrino to oscillate with the same frequency.}.
The results of this study can be immediately applied to the phases of neutrino emission for which collective effects are negligible, such as the neutronization  and/or late cooling phases. We use the results of this section later for comparison with  the results when $\nu\nu$ interactions are included.

\subsection{Effective width of the neutrino  emission region}

We will consider 
electron neutrinos and antineutrinos only and  
ignore complications related to the presence of  muon and tau neutrinos. 
Production of $\nu_e$ and $\bar{\nu}_e$  is dominated by the nucleon Urca processes:  
\begin{equation}
n + e^+ \rightarrow p + \bar\nu_e,~~~~~~p + e^- \rightarrow n + \nu_e. 
\label{eq:urca}
\end{equation}
The number of neutrinos  emitted from a unit size volume at a  given point (in unit of time) 
is given by the emissivity $J$:  
\begin{equation}
  \label{eq:Jnucleon}
  J_{\rm nucleon}(E,r) = \frac{1}{2}\sigma_{\rm nucleon} n_N(r) n_{e^\pm}(E,r),
\end{equation}
where $n_N$ is the number density of nucleons and $n_{e^\pm}(E,r)$ 
is the number density of electrons or positrons 
with energy $E$ \footnote{The factor of $\frac{1}{2}$ arise because only left-handed electrons
feel the weak interaction.}. 
We approximate the electron number density using the  Boltzmann factor: 
\begin{equation}
  \label{eq:ne}
  n_{e^\pm}(E,r) = \frac{n_{e^\pm}^{\rm tot}(r)}{T(r)} e^{-E/T(r)},
\end{equation}
where  $T$ is the temperature of the medium  and 
$n_{e^\pm}^{\rm tot}(r)$ are the total number densities of electrons and positrons. 
Their difference is fixed by the electric charge neutrality of the medium if the density of protons is known.
The neutrino-nucleon cross-section (\ref{eq:urca}) equals approximately~\cite{Reddy:1997yr}
\begin{equation}
  \label{eq:Gammanucleon}
  \sigma_{{\rm nucleon}} = \frac{G_F^2}{\pi} (g_V^2 + 3 g_A^2) E^2,
\end{equation}
with $g_A = 1$ and $g_V = 1.23$. 

The probability that a neutrino 
travels from the production point $r_e$ to a given point $r$  
(and we consider radial motion here) without absorption equals  
\begin{equation}
  \label{eq:Notsc}
  p_{\rm survive}(r,r_e)  = \exp\left(-\int_{r_e}^r \sigma_{\rm nucleon} n_N(r') dr'\right). 
\end{equation}
Consequently,  the neutrino flux at radius $r$ is
\begin{equation}
  \label{eq:flux}
 F(E,r,\vec{v}) = \frac{dF(E,r)}{d\Omega} = 
\frac{1}{4\pi}\int_0^r  J_{\rm nucleon}(E,r_e) p_{\rm survive}(r,r_e) dr_e,
\end{equation}
where $d\Omega$ is the  solid angle around the direction determined by the neutrino velocity $\vec{v}$. 
We use this one dimensional picture for simplicity. In a realistic supernova, neutrinos are emitted in different directions. For non-radial motion, the effects of averaging are expected to be even stronger since neutrinos will spend more time in the neutrino emission region (or equivalently will have a wider production region).

\begin{figure}[tbp]
  \centering
  \includegraphics[width=0.8\textwidth]{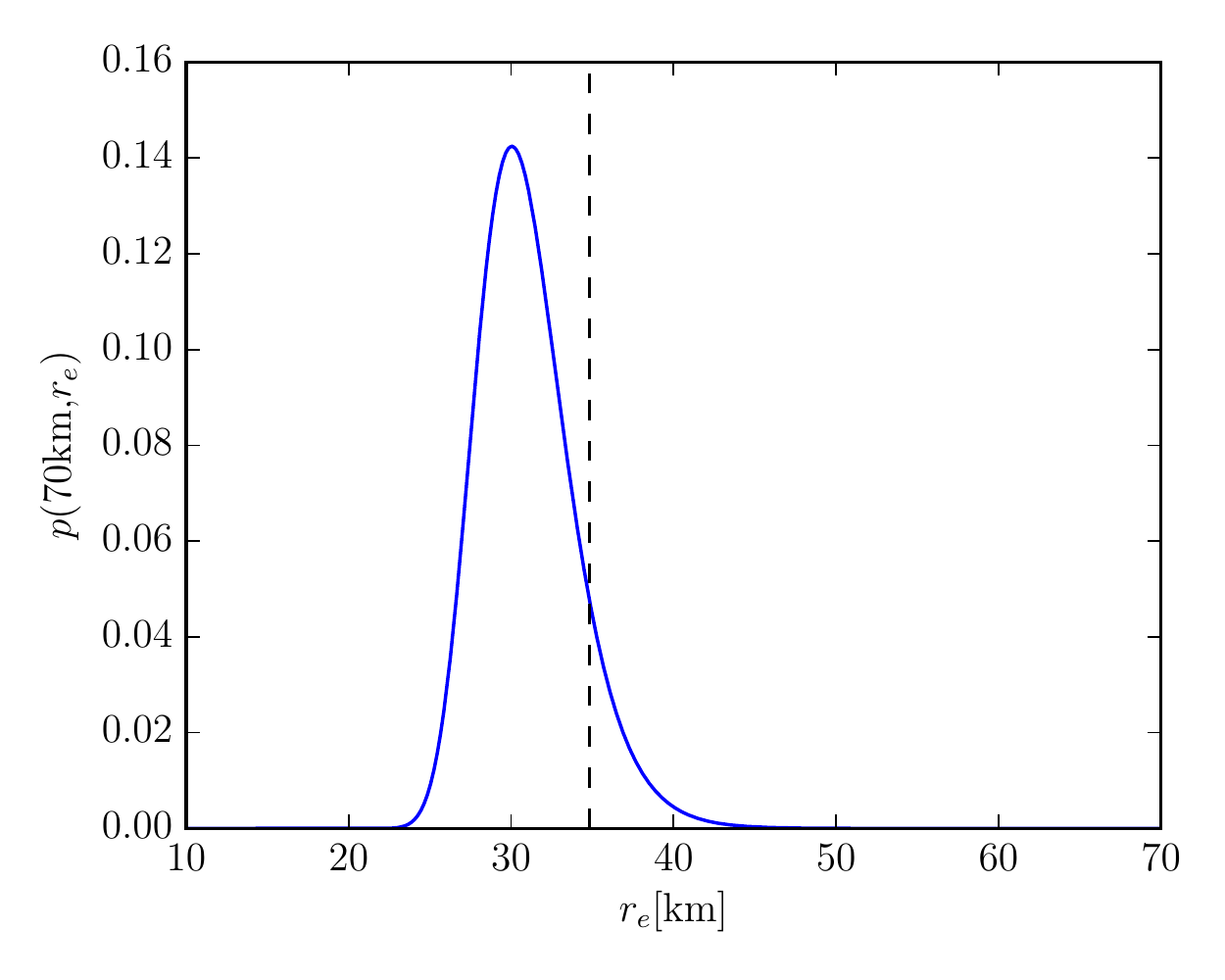}%
  \caption{The probability that neutrino   
with energy 10 MeV observed at the distance $r = 70$ km 
from the centre of a star is emitted at the distance $r_e$. 
We use the density distribution from  \eref{eq:density}, 
the electron fraction $Y_e = 0.4$ and the temperature profile in \eref{eq:temp}.
The dashed line indicates the position of the infinitesimal neutrino sphere according to \eref{eq:Rnusphere}. 
}
\label{fig:prob}
\end{figure}

Let us introduce  the probability $p(r,r_e)$ that a neutrino observed at $r$ has been emitted at $r_e$. 
According to (\ref{eq:flux}) 
\begin{equation}
  \label{eq:Pemm}
 p(r,r_e) = \frac{1}{4\pi F(E,r)}J_{\rm nucleon}(E,r_e) p_{\rm survive}(r,r_e).
\end{equation}
As an example,  we show the dependence of $p(r,r_e)$   
on $r_e$ in \fref{fig:prob} for large $r$ with
the density profile~\cite{Liebendoerfer:2003es}
\begin{equation}
  \label{eq:density}
  \rho(r) = \rho_0 e^{-r/r_0},
\end{equation}
where $\rho_0=3\times 10^{15}{\rm g}/{\rm cm}^3$ and $r_0=4$ km, and with
the temperature profile~\cite{Keil:2002in}
\begin{equation}
  \label{eq:temp}
  T = T_0 \frac{r_0}{r}, 
\end{equation}
where $T_0 = 50 $ MeV. 
The infinitesimal neutrino sphere is indicated by the dashed line. 
The dependence of $p$ on $r_e$ has character of a peak at a certain radius $r_{\rm max}$:
For $r_e > r_{\rm max}$, $p$ decreases because the nucleon density decreases. 
At $r_e < r_{\rm max}$, the absorption becomes strong and emitted neutrinos cannot escape.
According to \fref{fig:prob}, the width of the probability distribution  is $r_{\rm eff} \sim$ 5 - 6 km. 
For different  density and temperature profiles the width can vary by a factor 2-3.  

We use single energy neutrino fluxes which to some extend correspond to synchronised oscillations. So, the results for the synchronised oscillations are expected to resemble what we find in this section.

\subsection{Effect of averaging over the neutrino production region}
\label{sec:av}

We consider the system of two mixed neutrinos  $\nu_f = (\nu_e, \nu_x)$, where $\nu_x$ 
is a linear combination of $\nu_\mu$ and $\nu_\tau$. 
As the mass squared difference, we take  $\Delta m^2 = \Delta m_{31}^2$.
The wave function describing the neutrino flavour $\psi \equiv (\psi_e, \psi_\mu)^T$ 
evolves according to the equation 
\begin{equation}
  \label{eq:eom}
  i \partial_r \psi(r) = H(r) \psi(r) 
\end{equation}
with the Hamiltonian 
\begin{equation}
  \label{eq:Hdef}
  H(r) = \frac{1}{2}
  \begin{pmatrix}
     - \omega \cos 2\theta + V_e(r) & \omega \sin 2\theta \\ 
     \omega\sin 2\theta & \omega \cos 2\theta - V_e(r)
  \end{pmatrix}.
\end{equation}
Here $\omega \equiv \Delta m^2/2E$, $\theta$ is the vacuum mixing angle, 
and $V_e(r) = \sqrt{2} G_F (n^{\rm tot}_{e^-}(r) - n^{\rm tot}_{e^+}(r) )$ 
is the standard matter potential given by the  forward neutrino scattering on electrons and positrons. 

We introduce the instantaneous eigenstates of the Hamiltonian 
$\nu' \equiv (\nu_1', ~ \nu_2')^T$ which diagonalise  $H(r)$:
\begin{equation}
\nu_f =  U(\theta_m) \nu', ~~~~  U(\theta_m)^{\dagger} H   U(\theta_m) = H', 
\label{eq:mixing}
\end{equation}
where  $U(\theta_m)$ is the unitary mixing matrix and  $\theta_m(r)$  is the mixing angle in matter:
\begin{equation} 
\sin 2 \theta_m = \frac{\sin 2 \theta}{\sqrt{(\cos 2\theta - V_e/\omega)^2 + \sin^2 2\theta}} 
\approx \sin 2 \theta \frac{\omega}{V_e},  
\label{eq:mixangle}
\end{equation}
and 
\begin{equation}
  \label{eq:Hp}
 H'(r) = \frac{\omega_m(r)}{2}
  \begin{pmatrix}
     - 1 & 0 \\ 
     0 & 1 
  \end{pmatrix} 
\end{equation}
is the Hamiltonian  in the basis of instantaneous eigenstates
with $\omega_m(r)$ being the difference of the eigenvalues or the oscillation frequency:  
$$
\omega_m(r)  = \omega \sqrt{(\cos 2\theta - V_e/\omega)^2 + \sin^2 2\theta}. 
$$
In central parts of the supernova, the adiabaticity condition,  
$$
\gamma_{ad} \equiv  \frac{1}{\omega_m(r)} 
\frac{d \theta_m(r)}{dr} \ll 1, 
$$ 
is well satisfied. E.g., for the density profile in \eref{eq:density}, we find  
$\gamma_{ad} = 10^{-17}$ and $\gamma_{ad} =  10^{-8}$ 
for $r=50$ and $200$ km respectively.
Therefore, there are no transitions between the eigenstates, and they evolve independently: 
\begin{equation}
\psi_1' \propto \exp \left[\frac{1}{2} i \int \omega_m(r) dr \right], ~~~
\psi_2' \propto \exp \left[-\frac{1}{2} i\int \omega_m(r) dr \right]. 
\label{eq:eigev}
\end{equation} 
In the basis  of eigenstates, the electron neutrino is described by 
\begin{equation}
  \label{eq:psie}
  \psi_e(r) =
  \begin{pmatrix}
    \cos \theta_m(r) \\ \sin \theta_m(r)
  \end{pmatrix}.
\end{equation}
Then the evolution of a state produced as $\nu_e$ is, according to (\ref{eq:eigev}), given by
\begin{equation}
  \psi_e (r) =
  \begin{pmatrix}
\cos \theta_m(r_e) e^{0.5 i\int \omega_m(r) dr}  \\ 
\sin \theta_m(r_e) e^{- 0.5 i\int \omega_m(r) dr} 
  \end{pmatrix},
\label{eq:nue-evol}
\end{equation}
where the  $\theta_m(r_e)$ is the mixing angle in the production point. 

Let us introduce the correlation of the 
eigenstates in the state produced as $\nu_e$:  
\begin{equation}
  \label{eq:rho12}
  \rho_{12}'(r) = \psi'_1(r) \psi'^{*}_2(r) .
\end{equation}
It is this correlation given by the off-diagonal element of the density matrix
that is responsible for flavour conversion. 
According to (\ref{eq:nue-evol}) 
\begin{equation}
  \label{eq:rho12sol}
  \rho_{12}'(r,r_e) = \frac{1}{2}\sin 2\theta_m(r_e) \exp\left(i \int_{r_e}^r \omega_m(r') dr'\right).
\end{equation}
The  correlation averaged over the neutrino production region equals
\begin{equation}
  \label{eq:rho12av}
  \left<\rho_{12}'(r)\right> = \int_0^r p(r,r_e) \frac{1}{2}\sin 2\theta_m(r_e) 
\exp\left(i \int_{r_e}^r \omega_m(r') dr'\right)  dr_e , 
\end{equation}
where the emission probability  $p(r,r_e)$  is determined  in \eref{eq:Pemm}. 
From \eref{eq:rho12av}, we can immediately see two properties of the averaged correlation: 
\begin{enumerate}
\item The exponent introduces very fast oscillations, 
and this results in a strong suppression of the integral 
over $r_e$ (as it was noticed in \cite{Hansen:2018apu}). 
\item The emission probability  $p(r,r_e)$ decreases with $J \propto n_e n_N$ 
at large radii while $\sin 2\theta_m$ 
increases as $1/V_e \sim 1/n_e$ (for $V_e \gg \omega$) partially cancelling the dependence 
on $r_e$ in the integrand. 
Although the dependence on $n_{e^-}^{\rm tot}$ cancels, there is still a decrease of the integrand 
with $r_e$ due to the Boltzmann factor and the number density of nucleons. 
\end{enumerate}
For the exponential density profile in \eref{eq:density} and the temperature profile 
in \eref{eq:temp}, the averaging in \eref{eq:rho12av} can be performed analytically 
when the Boltzmann suppression of $n_e(E,r)$ is neglected. For $r \gg r_0$  
using that $\sigma_{\rm nucleon} n_N(r_0) r_0 \gg 1$, we obtain 
\begin{equation}
  \label{eq:rho12avana}
  \left<\rho_{12}'(r)\right> \approx - i \frac{1}{2} \sin 2\theta_m \frac{1}{ r_0 V_e} 
\frac{r_0^2}{ l_{\rm abs}^2 } ,
\end{equation}
where $l_{\rm abs} \equiv (\sigma_{\rm nucleon} n_n)^{-1}$ is the neutrino absorption length and $r_0$ is defined below \eref{eq:density}. 
(The oscillatory terms in $\left<\rho_{12}'(r)\right>$ are suppressed by $e^{-\sigma n_0 r_0}$.)
According to (\ref{eq:rho12avana}), apart from  very small $\sin 2\theta_m $,
the correlation contains an additional  suppressed  factor 
$r_{\rm eff} V_e$ where $r_{\rm eff} = l_{\rm abs}^2/r_0$. 
The effective width  of the production region is determined by the absorption length,  
so that averaging becomes weaker with the decrease of $l_{\rm abs}$. 
Numerically, $l_{\rm abs} \sim 5$ km for $n_N \sim 2 \times 10^{35} {\rm cm}^{-3}$ which 
corresponds to $r \sim 35$ km for the density profile in \eref{eq:density}. The dependence 
on radius in $\sin 2\theta_m$, $V_e$, and $l_{\rm abs}$ cancels in \eref{eq:rho12avana}, so 
these three quantities can be evaluated at any radius as long as it is the same for all three.

\begin{figure}[tbp]
  \centering
  \includegraphics[width=0.8\textwidth]{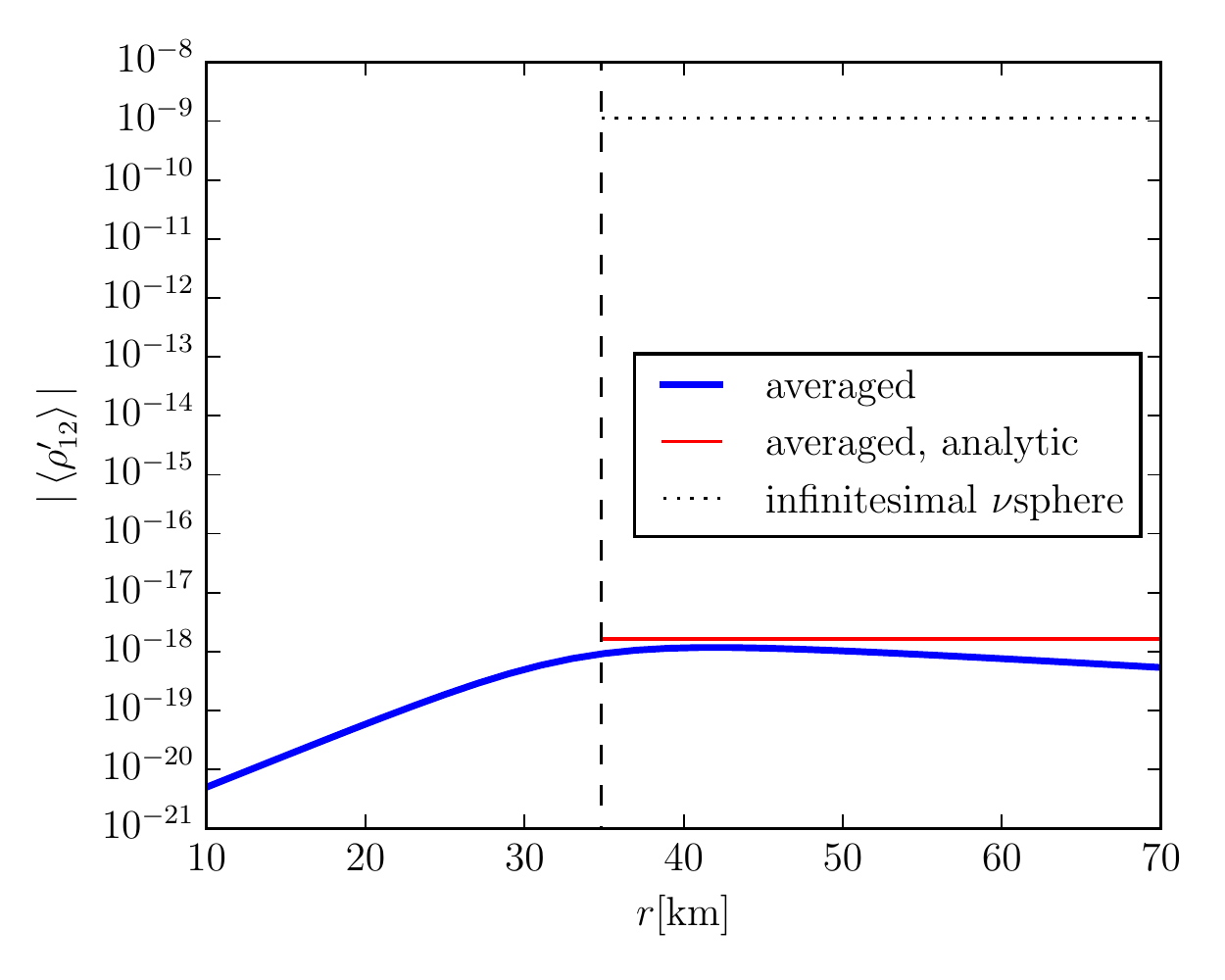}
  \caption{The averaged correlation in the eigenstates basis as function of the distance 
from the centre of the star. Shown are results of the numerical integration in  
\eref{eq:rho12av} (blue line) and the analytic result in \eref{eq:rho12avana} (red line).  
In the calculations, the emission probability of \fref{fig:prob} is used. 
The neutrino energy is 10 MeV and the electron fraction is assumed to be $Y_e=0.4$. 
For comparison we show the result for a neutrino sphere
of infinitesimal width at $r_e = 35$ km as the dotted line.}
  \label{fig:rhoav}
\end{figure}

The results of the numerical evaluation of the averaged correlation in \eref{eq:rho12av} 
is shown in \fref{fig:rhoav} together with the analytical estimation 
from \eref{eq:rho12avana}.  
The vertical dashed line indicates the position of the infinitesimal neutrino sphere 
from \eref{eq:Rnusphere}. The difference between the numerical and analytic result  
is due to the neglected Boltzmann suppression in the analytic calculation. 
Notice that the initial increase of $|\left< \rho'_{12}\right>|$ 
is due to contributions from neutrinos emitted at the  largest $r_e$ (smallest density)
since the effective mixing angle increases with the decrease of density. 

We see that the  integration over the production region 
leads to a suppression  of the off-diagonal term in the density matrix by $\sim 10$ orders of magnitude 
with respect to the result for an infinitesimal neutrino sphere.  
Similar suppression of the order $1/V_e  r_{\rm eff}$ is realised for steeper 
density profiles with $R_{\nu \rm sphere} = (10-20)$ km. 
This suppression factor should be included in  the initial condition for further collective oscillations. 
We will justify the factorisation of  averaging in the production region and collective oscillations 
outside the production region later using a solvable example.

\section{Averaging in the presence of $\nu \nu$ interactions}
\label{sec:simple}

\subsection{Model with $\nu \nu$ interactions}

In order to  understand how the emission of neutrinos from different 
points influences collective oscillations, 
 we will consider the simple system with neutrino self-interactions, 
two intersecting fluxes of collinear neutrinos moving in the plane $(x, z)$~\cite{Raffelt:2013isa, Mangano:2014zda, Hansen:2014paa, Duan:2014gfa, Mirizzi:2015fva, Martin:2019kgi}. 
The fluxes have angles $\beta$ (right moving) and $\pi - \beta$ (left moving)
with respect to the $x$-axis (the same angles with respect to $z$).
Both fluxes contain equal number densities of
neutrinos and antineutrinos and all neutrinos have the same
vacuum frequency $\omega$. The fluxes are moving in a uniform medium with matter potential $V_e$. 
We consider the production of electron (anti)neutrinos only.

The set-up is symmetric with respect to the reflection 
$\beta \rightarrow \pi - \beta$, and therefore left and right moving 
neutrinos have the same evolution. 

We took equal numbers of neutrinos and antineutrinos for simplicity. This is the most favourable case for collective oscillations where bi-polar oscillations can start immediately without a delay.
We assume that the neutrinos are produced  in a region of width $\Delta z$,  
so that the coordinates of production points $z_e$ are in the interval $0 \leq z_e \leq \Delta z$. 
In this model the time (distance) of propagation along the trajectory is uniquely related 
to the coordinate $z$: 
$$
t = t(z) = \frac{z}{s_\beta}, ~~~dt = \frac{1}{s_\beta} dz,~~~~ s_\beta \equiv \sin \beta,
$$
and in what follows, we will use $z$ to describe the evolution. 

We will treat the problem in terms of density matrices $\rho(z_e, z)$ and $\bar{\rho}(z_e,z)$
which describe the state of  neutrinos and antineutrinos
produced as $\nu_e$ and $\bar{\nu}_e$ at $z_e$ as a function of coordinate $z$.  
The evolution equation for  $\rho(z_e, z)$ in the flavour basis is
\begin{equation}
  \label{eq:simpleeom}
  i s_\beta\frac{d}{dz} \rho(z_e,z) = [H(z), \rho(z_e,z)],
\end{equation}
where   
\begin{equation}
 \label{eq:Hsimple}
  H(z) = \frac{\omega}{2}
  \begin{pmatrix}
    -\cos 2\theta & \sin 2\theta \\ \sin 2\theta & \cos 2\theta
  \end{pmatrix}
  + \frac{1}{2}
  \begin{pmatrix}
    V_e & 0 \\ 0 & -V_e
  \end{pmatrix}
  + \mu \int_0^{z} \left[\rho(z_e',z) - \bar{\rho}^*(z_e',z)\right] p_{\nu}(z_e')dz_e'.
\end{equation}
Here $p_\nu(z_e')$ is the probability that a given free streaming neutrino is produced at the point $z_e'$ and $\mu = \sqrt{2} G_F n_\nu$, where $n_\nu = n_{\nu_e} - n_{\nu_x}$ is the effective density of neutrinos at $z>\Delta z$,  
so $\mu$ is the total potential produced by the neutrino background outside the production region. 
The first and second terms in \eref{eq:Hsimple} coincide with the Hamiltonian in (\ref{eq:Hdef}), 
the last term is due to $\nu\nu$ interactions.
We take equal numbers of neutrinos and antineutrinos,   
so that ${\rm Tr}(\rho) = {\rm Tr}(\bar{\rho}) = 1$. 
We assume the hierarchy of frequencies 
$$
V_e \gg \mu \gg |\omega|
$$
typical for the central parts of supernovae. 

In the Hamiltonian for antineutrinos,
$V_e$  has the opposite sign, and  $\rho$ and $\bar \rho$ are swapped.
The initial condition reads $\rho_{ee} =  \bar{\rho}_{ee} = 1$. 

For simplicity, we assume  that neutrinos are produced 
in the layer $z \in [0,\Delta z]$ uniformly,  so that the emission probability introduced in the 
previous section equals
\begin{equation}
  \label{eq:psimple}
  p_{\nu}(z_e) =
 \begin{cases}
    \frac{1}{\Delta z} &{\rm for} \quad 0 < z_e < \Delta z, \\ 0 & {\rm otherwise}
  \end{cases}. 
\end{equation}
In contrast to the previous section, we impose the size $\Delta z$ by hand.

With the emission profile (\ref{eq:psimple}) and a constant matter density, the expression  
\eref{eq:rho12av}, where $\nu \nu$ interactions are not accounted for, is simplified for 
$z>\Delta z$: 
\begin{equation}
  \label{eq:rho12avsimple}
  \left<\rho_{12}'(z)\right> = \frac{1}{2 \Delta z}\sin 2\theta_m 
\int_0^{\Delta z} \exp\left[i \omega_m (z-z_e)/s_\beta \right]  dz_e 
\end{equation}
if  the neutrino background contribution is neglected in $\theta_m$ and $\omega_m$.  
Furthermore, since  $\mu, \omega \ll V_e$, we can neglect $\mu$ and $\omega$ in $\omega_m$,  
so that $\omega_m \approx V_e$ and integration in  
(\ref{eq:rho12avsimple}) can be performed explicitly giving 
\begin{equation}
  \label{eq:rho12avsimple1}
  |\left<\rho_{12}'(z)\right>| = \frac{s_\beta \sin 2\theta_m}{V_e \Delta z}
\end{equation}
for the phase $\pi + 2\pi k$. Thus, as expected, integration leads to additional suppression 
of the correlation by a factor $(V_e \Delta z)^{-1} \ll 1$. 

Throughout the emission region, the number of neutrinos increases as more neutrinos are emitted. 
The fraction of neutrinos that has been emitted  by a given  point z equals 
\begin{equation}
  \label{eq:nemitted}
  f_\nu(z) = \int_0^z p_{\nu}(z_e) dz_e = 
  \begin{cases}
    \frac{z}{\Delta z} & {\rm for} \quad 0 < z < \Delta z , \\ 1 & {\rm for} \quad z > \Delta z.
  \end{cases}
\end{equation}
Then the potential due to the neutrino background can be parametrised as 
\begin{equation}
  \label{eq:mutildedef}
    \tilde{\mu}(z)  = \mu f_\nu(z)
\end{equation}
which is the fraction of the potential due to $\nu \nu$ interactions accumulated to a given point $z$. 
For our model of emission (\ref{eq:nemitted}):
\begin{equation}
\tilde{\mu}(z) \equiv  
 \begin{cases}
 \mu \frac{z}{ \Delta z}  
     & {\rm for} \quad 0 < z < \Delta z , \\ 
\mu  & {\rm for} \quad z > \Delta z.
  \end{cases}
\label{eq:mugrow}
\end{equation} 

We are interested in the evolution of flavour of the entire ensemble of neutrinos.  
Therefore, we define the total normalised density matrix as  
\begin{equation}
  \label{eq:rhoint}
  \rho^{\rm int}(z) = \frac{1}{f_\nu (z)}\int_0^z \rho(z_e, z) dz_e .
\end{equation}
The density matrix $\rho^{\rm int}$ is normalised in the same way as a single state. 
So in what follows, we will talk about averaging of the density matrix over the production region. 

\subsection{Results of numerical computations}

We will study averaging effects depending on the width of the production region 
$\Delta z$.  Since the total number of emitted neutrinos $\mu$ is fixed, 
the emissivity decreases with the increase of $\Delta z$. 
To see  10 orders of magnitude suppression 
in a numerical computations, about $\mathcal{O}(10^{10})$ different emission 
points have to be included, which is unfeasible. 
Therefore, we use a much weaker hierarchy between the different frequencies involved and 
assume a size of the emission region which is at most a few orders 
of magnitude larger than the oscillation length.  
We take the inverted hierarchy, $\omega < 0$, 
and for definiteness the following values of parameters:
\begin{equation}
V_e = 30 |\omega|, ~~~ \mu = 3 |\omega|,  ~~~ \theta = 0.15, ~~~~ \beta = \pi/4. 
\label{eq:parameters}
\end{equation} 
We solve the evolution equations numerically to obtain $\rho(z_e,z)$ and $\rho^{\rm int}$. 
Details of our computations are given in  Appendix \ref{sec:numerical}.
(Although   $s_\beta = 1/\sqrt{2}$, we will still keep $s_\beta$ in formulas below.)

\begin{figure}[tbp]
  \centering
  \includegraphics[width=0.7\textwidth]{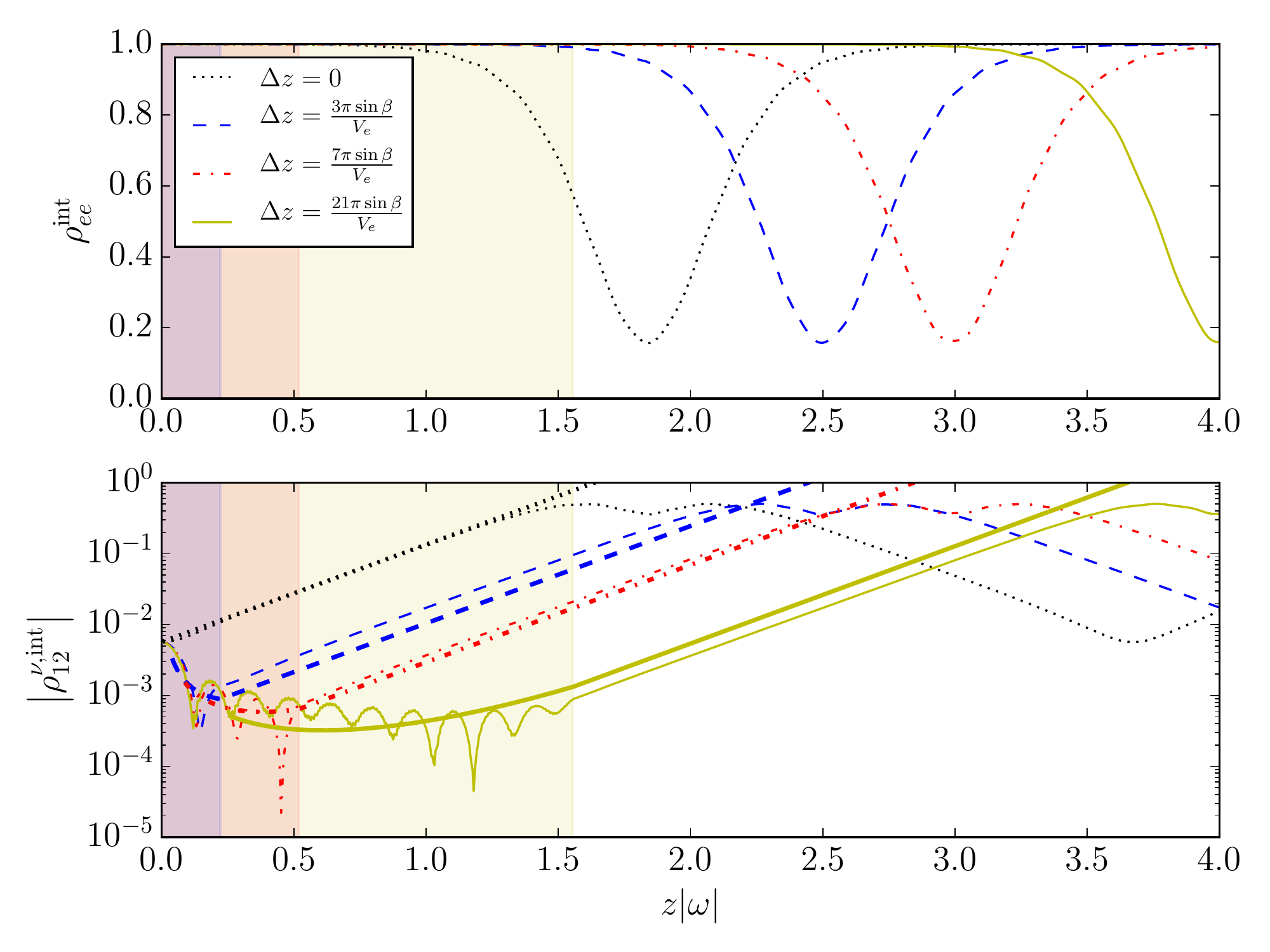}%
  \caption{The elements of density matrices averaged over narrow production regions as functions 
of the distance from the beginning of the region for different values of width $\Delta z$ (shadowed). 
{\it Upper panel:} The $\nu_e$ survival probability ($\rho_{ee}$). 
{\it Bottom panel:} The correlation -- the off-diagonal term of the density matrix 
in the eigenstates basis $\rho_{12}$. Thick lines show analytic results based on the solution of the linearised equations.}
  \label{fig:beam_lown}
\end{figure}

\begin{figure}[btp]
  \centering
  \includegraphics[width=0.8\textwidth]{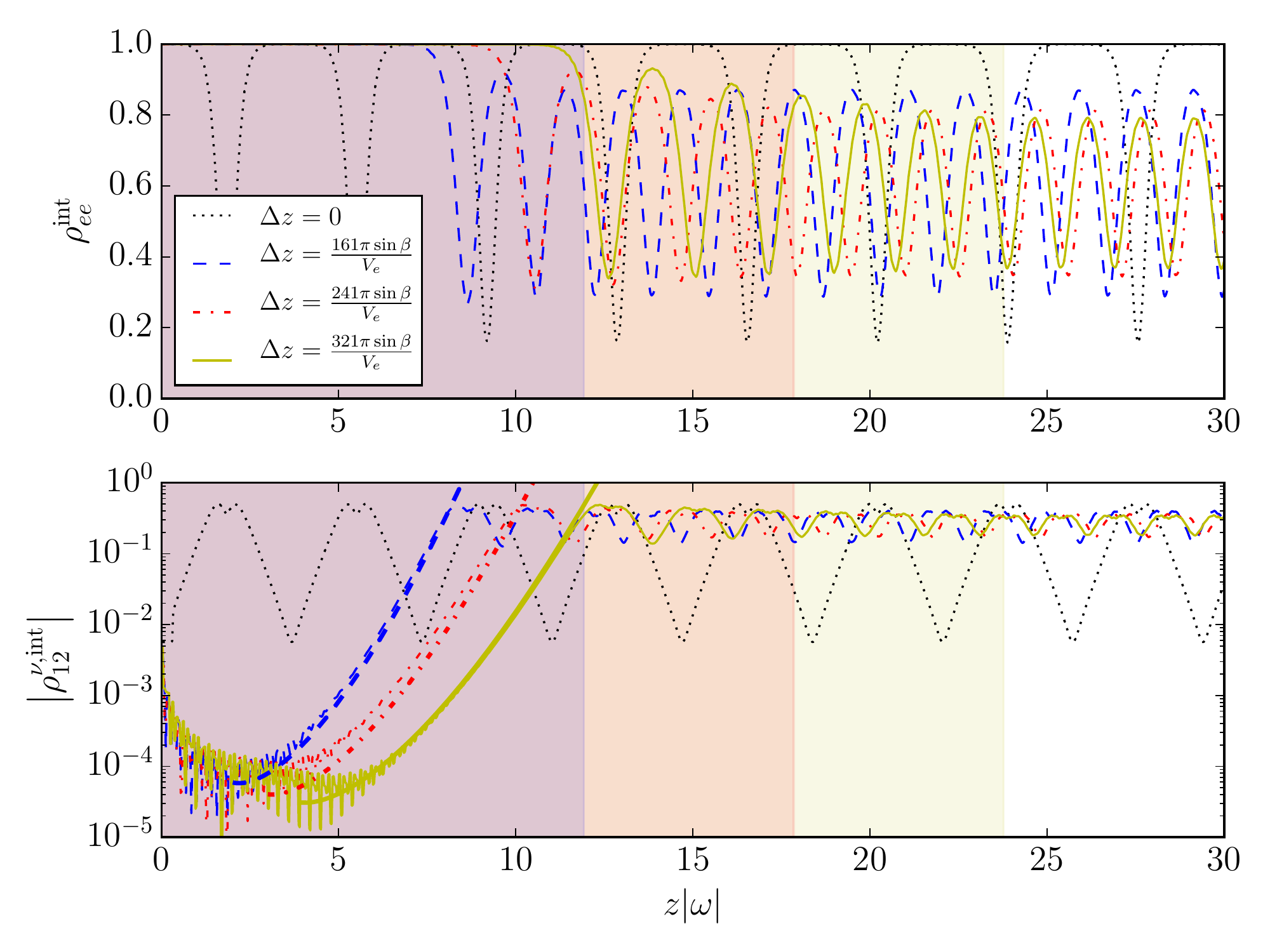}%
  \caption{The same as in \fref{fig:beam_lown}, but for wide production regions.}
\label{fig:beam_highn}
\end{figure}

The result of averaging depends on the exact value of $\Delta z$. For definiteness,  
we will take $\Delta z$  to be semi-integer numbers of matter refraction lengths  
$2\pi/V_e$: $\Delta z = (2n+1) \pi s_\beta/V_e$, 
where $n \geq 0$ is an integer.
Due to the presence of neutrinos in the background, this does not correspond exactly to 
a semi-integer number of oscillations.
Numerically, for values of parameters 
(\ref{eq:parameters}) we have  $\Delta z |\omega| 
 = 0.074 (2n+1)$.  
The results  of computations are shown in  \fref{fig:beam_lown} for relatively narrow 
production regions: $n=1,~ 3,~ 10$, which corresponds to 
$\Delta z |\omega| < 1.5$, 
 and  in \fref{fig:beam_highn} 
for the wider regions with $n=80,~ 120,~160$ ($\Delta z |\omega| > 12$). For comparison, 
we also present results for a surface emission at $z = 0$ with  $\Delta z = 0$. 

The upper panels show, as expected, a gradual delay of the oscillations of $\rho_{ee}$ with an increase of the emission region. $\rho_{ee} \approx 1$ in the initial phase, and then the evolution proceeds with bi-polar oscillations.  
In \fref{fig:beam_lown}, the strong transition effect develops outside the 
production region. Consequently, the depth and the period of oscillations do not depend 
on $\Delta z$. The only effect of the averaging is a delay of the strong transition -- 
a shift of the first minimum. 
The exponential growth is the same for all $\Delta z$ at $z > \Delta z$. 
For large values of $\Delta z$ ($n=80$, $120$, and $160$),
strong conversion develops already within the emission region.

%

Although initially, $\rho_{ee} \approx 1$, and no change is seen, 
important dynamics  occurs in this initial phase 
which eventually leads to the delay of visible bi-polar oscillations. 
This can be seen and understood in detail in terms of the correlation  of the eigenstates 
$\rho_{12}$ (\ref{eq:rho12}). 

Eqs. (\ref{eq:simpleeom}) and (\ref{eq:Hsimple}) and the corresponding density matrices
are given in the flavour basis. To explore 
the correlations $\rho_{12}$, we need to make a transformation to the eigenstate basis.  
The relation between the off-diagonal elements 
of the density matrices in the flavour and eigenstates bases 
(see Appendix \ref{sec:transform}) is given by  
\begin{equation}
\label{eq:rhonutilde}
\left|\rho_{12}^{\nu,{\rm int}}(z) \right| \approx  \left| 
\rho_{ex}^{{\rm int}} (z) - \frac{\omega \sin 2\theta}{2(V_e - \omega \cos 2\theta)}\left(1 + 2\frac{\tilde{\mu}(z)}{V_e}\right) 
\right| , 
\end{equation}
where $\tilde{\mu}(z)$ is defined in (\ref{eq:mugrow}). 
Thus, the transition to the eigenstate basis adds a small term which becomes
negligible in the period  of strong collective effects. 
So,  apart from the initial phase of evolution, when
$\rho_{ex} \approx 0 $, we have 
\begin{equation}
|\rho^\nu_{12}|  \approx  |\rho_{ex}|,   ~~~~|\rho_{ex}| = \sqrt{\rho_{ee}(1 - \rho_{ee})}. 
\label{eq:relafe}
\end{equation}

The results of evolution of $\rho_{12}^{\nu, \rm int}$  are shown in the lower panels of  
\fref{fig:beam_lown} and \fref{fig:beam_highn}.
According to the figures (bottom panel), there are two benchmark points: 

\begin{itemize}

\item $z_0$ -- The point  of onset of the exponential growth of the correlation due to
collective effects.
 
\item $z_1$ -- The  coordinate at which  maximum  of the correlation defined by the condition
$|\langle \rho_{12}^{\nu, \rm int} (z_1) \rangle| \sim 0.5$ is achieved for the first time.

\end{itemize}
Correspondingly, one can identify three phases of the evolution: 
\begin{itemize}
\item $z \in [0, z_0]$ --  The  averaging phase 
when  the correlation $|\langle \rho_{12}^{\nu, \rm int} \rangle |$ 
decreases due to integration over the production region
and collective effects can be neglected; 

\item $z \in [z_0, z_1]$ --   Nearly exponential growth of the 
correlation $|\langle \rho_{12}^{\nu,\rm int} \rangle |$; here 
collective effects start to dominate.

\item $z >  z_1$ --  Regular (or quasi-regular) bi-polar oscillations. 
\end{itemize}

The difference of frequencies for different sizes of the production region that can be seen in \fref{fig:beam_highn} is related to a difference in neutrino densities since the density equals $n^{\rm tot} z/\Delta z$ in our model, and the total number density of neutrinos outside the production region, $n^{\rm tot}$, is the same for all widths. So the bigger the production region, the smaller the neutrino density, and the smaller the frequency as can be seen in \fref{fig:beam_highn} (upper panel). Outside the production regions, all frequencies are equal. (The only exception is $\Delta z = 0$ for which the flavour evolution takes the system very close to the 'fixed point' at $\rho^{\nu}_{12} = 0$ ($\rho_{ee} \approx 1$), where the flavour only changes slowly. This delays the oscillations and gives a significantly lower frequency.)

In our model with increasing $\tilde{\mu}$ in the emission region,  
the onset of collective effects is always in the 
emission region: $z_0 \leq \Delta z$. 
Depending on the size of $\Delta z$ with respect to $z_1$, 
there are two different situations:

\begin{enumerate}
\item Narrow production region: $\Delta z < z_1$. 
The instability develops partly outside the production region (\fref{fig:beam_lown}). 
In this case for $z > z_1$, oscillations proceed with constant parameters (depth and period). 
Furthermore, $\rho_{ee} = 1$ at the maxima. So the asymptotic behaviour starts from $z_1$. 

If the production region is very narrow, $\Delta z \ll z_1$, we can neglect 
effect of collective oscillations in the emission region and 
consider the exponential growth only outside this region with certain 
boundary conditions. The description is further simplified in this case.  
That would correspond to the  situation described in \sref{sec:effwidth}. 

\item Wide production region: $\Delta z > z_1$. In this case the instability (with exponential growth) 
develops completely inside the production region (see \fref{fig:beam_highn}). 
In the interval after the development of the instability between $z_1$ and $\Delta z$,
bi-polar oscillations proceed with decreasing 
depth, and asymptotic oscillations start from $z_{\rm as} =  \Delta z$. 

The off-diagonal term in the density matrix peaks twice before it starts to decrease again. 
The maximal value of $\rho^\nu_{12}$ 
(maximal possible correlation between the two eigenstates) is achieved 
when $\rho_{ee}^{\rm int}\sim 0.5$, as can be seen from \eref{eq:relafe}.  
\end{enumerate}

In what follows, we will consider these two possibilities separately. 
Notice that for the adiabatic propagation, $|\rho^\nu_{12}| =$ constant, and therefore 
changes of  $|\rho^\nu_{12}|$ reflect adiabaticity violation or new neutrinos joining the system.

\section{Analytic consideration using linearised equations}
\label{sec:linear}

As it follows from Fig. \ref{fig:beam_lown},  the exponential growth of the off-diagonal  
elements of $\rho^\nu$ with $z$  starts when $\rho^\nu_{12} \ll 1$ and proceeds 
until $\rho^\nu_{12} \sim 0.3$. In this range of $z$, one can use linearised 
equations of evolution which allow us to solve the problem analytically. 
The rate of exponential growth of $\rho^\nu_{12}$  can be calculated through 
linear stability analysis~\cite{Banerjee:2011fj}. The analysis is based on 
linearised evolution  equations for the off-diagonal part of the density matrix. 

The linear analysis has to be performed around a 'fixed point' -- that is a point in 
the space of density matrices 
where $d\rho/dz=0$ to zeroth order in the small off-diagonal element, and evolution 
only starts at linear or higher order. 
We find such a fixed point in the basis of propagating states where a vanishing 
off-diagonal term in the density matrix corresponds 
to independently propagating states with no oscillations.

We consider density matrices of neutrinos and antineutrinos produced as $\nu_e$ and $\bar{\nu}_e$ at $z_e$. 
In the case of adiabatic evolution, the basis of propagating states approximately coincide
with the eigenstate basis -- the basis of eigenstates of the instantaneous Hamiltonian  
where the $\nu \nu$ interaction term is included. For the calculations, we use the 
eigenstate basis (see Appendix \ref{sec:eigenstate} for details).
Since the matrices have trace 1, we can present them in the eigenstate basis as 
\begin{equation}
  \label{eq:defS}
  \begin{aligned}
  \rho^\nu(z_e,z) &= \frac{1}{2}
  \begin{pmatrix}
    1 & 0 \\ 0 & 1
  \end{pmatrix}
  +\frac{1}{2}
  \begin{pmatrix}
    -s(z_e,z) & -S^*(z_e,z) \\ -S(z_e,z) & s(z_e,z)
  \end{pmatrix}
  ,\\
  \bar{\rho}^{\nu}(z_e,z) &=  \frac{1}{2}
  \begin{pmatrix}
    1 & 0 \\ 0 & 1
  \end{pmatrix}
  +\frac{1}{2}
  \begin{pmatrix}
    \bar{s}(z_e,z) & \bar{S}^*(z_e,z) \\ \bar{S}(z_e,z) & -\bar{s}(z_e,z)
  \end{pmatrix}
  ,
  \end{aligned}
\end{equation}
for  $0<z_e<\Delta z$. For  $z_e > \Delta z$,  $\rho^{\nu}(z_e,z) = \bar{\rho}^{\nu}(z_e,z) = 0$.
The traces play no role in the evolution equation,
and furthermore, $s^2 + S^2 = 1$. 
In the limit $V_e \gg \omega$ in the neutrino emission region,  
we have $S \ll 1$, and therefore $s = 1$ ($\bar{s}= 1$) to first order in $S$ ($\bar S$). 
At the production point $S = -\sin 2\theta_\nu$ and $\bar{S} = \sin 2\theta_{\bar{\nu}}$.

\subsection{Equations for the integrated $S$}
\label{sec:intf}

Using (\ref{eq:Hsimple}) and (\ref{eq:simpleeom}), we can derive the Hamiltonian  
and the evolution equation for the density matrix in the eigenstate basis (see Appendix~\ref{sec:eigenstate} for more details). 
Inserting the expressions from (\ref{eq:defS}), we get
the following equations to lowest order in $S$ and  $\bar{S}$ 
\begin{equation}
  \label{eq:linSgen}
  \begin{aligned}
    i s_\beta \frac{d}{dz} S(z_e,z) &= \left(- \omega + V_e \right) S(z_e,z)  
+ \mu \int_0^{z} p_{\nu}(z_e')  \left[ \bar{S}(z_e',z) - S(z_e',z) \right] dz_e', \\
    i s_\beta \frac{d}{dz} \bar{S} (z_e,z) &=  \left( \omega + V_e \right)  \bar{S}(z_e,z) + 
\mu \int_0^{z} p_{\nu}(z_e')  \left[ \bar{S}(z_e',z) - S(z_e',z) \right] dz_e'.
  \end{aligned}
\end{equation}
Eqs.~(\ref{eq:linSgen}) is a system of two coupled equations 
for the neutrino and antineutrino modes. 
Notice that the evolution equation for $\rho^\nu$ derived from (\ref{eq:simpleeom}) gives two connected equations 
for $S$ and $S^*$ which are equivalent, as a result of $\rho = \rho^{\dagger}$. 
The same  holds for $\bar\rho$. Consequently, we have only 
two equations in the neutrino-antineutrino system instead of four. 

We will search for a solution for the individual modes (produced at $z_e$) in the form 
\begin{equation}
  \label{eq:Ssol}
  \begin{pmatrix}
  S(z_e,z)\\\bar{S}(z_e,z)    
  \end{pmatrix}
  = \exp\left(-i s_\beta^{-1}\int_{z_e}^z \Omega(z') dz' \right) 
  \begin{pmatrix}
  Q(z_e) \\\bar{Q}(z_e)
  \end{pmatrix}
\end{equation}
with the integral in the exponent. 
Inserting  (\ref{eq:Ssol}) into (\ref{eq:linSgen}), we find
\begin{equation}
  \label{eq:linQ}
  \begin{aligned}
    \left[\Omega(z) + \omega - V_e \right] e^{-i s_\beta^{-1} \int_{z_e}^z \Omega(z') dz'}  Q(z_e) & 
= \mu \int_0^z p(z_e') e^{-i s_\beta^{-1} \int_{z_e'}^z \Omega(z'') dz''} \left[ \bar{Q}(z_e')- Q(z_e') \right] dz_e', \\
    \left[ \Omega(z) - \omega - V_e \right] e^{-i s_\beta^{-1} \int_{z_e}^z \Omega(z') dz'}  \bar{Q}(z_e) & 
= \mu \int_0^z p(z_e') e^{-i s_\beta^{-1} \int_{z_e'}^z \Omega(z'') dz''} \left[\bar{Q}(z_e') - Q(z_e') \right] dz_e'.
  \end{aligned}
\end{equation}
The integrated and normalised element $S^{\rm int}$ equals    
\begin{equation}
  \label{eq:Sint}
  S^{\rm int}(z) = 
\frac{1}{f_\nu (z)} \int_0^{z} p(z_e)  S(z_e, z) dz_e 
= \frac{1}{f_\nu (z)} \int_0^{z} p(z_e) e^{-i s_\beta^{-1} \int_{z_e}^z \Omega(z') dz'}  Q(z_e) dz_e,
\end{equation}
and a similar expression,  $\bar{S}^{\rm int}(z)$,  can be written for antineutrinos.
Multiplying Eqs.~(\ref{eq:linQ}) by $p(z_e)/ f_\nu(z)$ 
and integrating over $z_e$, 
we obtain equations for the integrated modes (\ref{eq:Sint}), which can be expressed as
\begin{equation}
  \label{eq:matrlin}
  \begin{pmatrix}
    -\omega + V_e - \tilde{\mu} - \Omega & \tilde{\mu} \\ - \tilde{\mu} & \omega + V_e + \tilde{\mu} - \Omega
  \end{pmatrix} 
\begin{pmatrix}
    S^{\rm int} \\ \bar{S}^{\rm int}
  \end{pmatrix}
 = 0,
\end{equation}
where $\tilde{\mu}(z)$  was defined in \eref{eq:mutildedef} and  (\ref{eq:mugrow}). 
A non-trivial solution of the system of linear equations in \eref{eq:matrlin}
for $S^{\rm int}$  exists if the  determinant 
of the matrix in  (\ref{eq:matrlin}) is zero. 
This gives the expression 
\begin{equation}
  \label{eq:Omegaint}
  \Omega(z) = V_e \pm \sqrt{\omega (2 \tilde{\mu}(z) + \omega)}. 
\end{equation}
The key feature of (\ref{eq:linSgen}) is that it is anti-symmetric 
and nontrivial evolution occurs due to couplings of neutrinos and antineutrinos 
given by $\mu$.  Without $\nu \nu$ interactions ($\tilde{\mu} = 0$), the equations decouple. 

The exponentially growing collective mode appears when ${\rm Im} (\Omega) > 0$. 
Since $\tilde{\mu}$ is  positive, this leads to the conditions 
\begin{equation}
2\tilde{\mu} > |\omega|, ~~~~{\rm for }~~~~  \omega < 0.  
\label{eq:instability}
\end{equation}
The equality $\tilde\mu(z_0) = |\omega|/2$ determines  the coordinate of the onset 
of collective oscillations: 
\begin{equation}
  \label{eq:onset}
  z_0 = \Delta z \frac{|\omega|}{2 \mu}.
\end{equation}
At this $z$, the  exponential growth of the correlation starts. 
Here we used the explicit expression for $\tilde\mu(z_0)$ (\ref{eq:mugrow}). 
In terms of $z_0$, we can write  $ \sqrt{|\omega| (2 \mu - |\omega|)}  = |\omega| \sqrt{z/z_0 - 1}$.
The normalisation factors $Q(z_e)$ and $\bar{Q}(z_e)$ are determined by the initial 
conditions $Q(z_e) = S(z_e, z = z_e) $ and $\bar{Q}(z_e) = \bar{S}(z_e, z = z_e) $. 
Since electron neutrinos are produced, we find that
\begin{equation}
  \label{eq:Qze}
  -Q(z_e) \approx \bar{Q}(z_e)  \approx 2 \rho^\nu_{12}(z_e) \approx  
\sin 2 \theta_m \approx 
\frac{\omega \sin2\theta}{V_e}, 
\end{equation}
and they  weakly depend on $z_e$. 
(The dependence arise due to $\tilde{\mu}(z)$ and it disappears if $\tilde{\mu}(z)$ is neglected in comparison to $V_e$.) 

Thus, the solution (\ref{eq:Sint}) equals approximately 
\begin{equation}
  \label{eq:Sint1}
  S^{\rm int}(z) = - \frac{\sin2\theta_m}{f_\nu(z)} 
\int_0^{z} p(z_e) e^{-i s_\beta^{-1} \int_{z_e}^z \Omega(z') dz'}  dz_e,
\end{equation}
where  $\Omega(z)$ is determined in (\ref{eq:Omegaint}). 
In terms of the onset parameter  $z_0$ from (\ref{eq:onset}), $\Omega(z)$  can be written 
as  
\begin{equation}
\label{eq:Omegaint2}
  \Omega(z) = V_e + \frac{|\omega|}{\sqrt{z_0}} 
\begin{cases}
(z_0 - z)^{1/2}  & {\rm for} \quad 0 < z < z_0,  \\
i (z - z_0)^{1/2}  & {\rm for} \quad z_0 < z < \Delta z, \\
i (\Delta z - z_0)^{1/2}  & {\rm for} \quad  z > \Delta z .
\end{cases}
\end{equation}
Integration of $\Omega$ in the exponent of (\ref{eq:Sint1}) can be done explicitly: 
\begin{equation}
\label{eq:Omegaint3}
\frac{1}{s_\beta}\int_{z_e}^z \Omega(z') dz' = 
\frac{1}{s_\beta} \left[V_e (z - z_e)  +  I(z, z_e) \right]. 
\end{equation}  
The integral  $I(z, z_e)$  splits into $z$- dependent and  $z_e$-dependent parts: 
\begin{equation}
\label{eq:Omegaint4}
I(z, z_e) = I (z) + I(z_e).   
\end{equation} 
Taking into account that $\Omega(z)$ has different expressions in different ranges of 
$z$, we find that there are four possibilities  depending on the relative values of 
$z$, $z_e$,  $z_0$ and $\Delta z$.  In all the cases we will take $z > z_0$. 

\begin{enumerate}
\item $z < \Delta z$ (wide emission region), $z_e < z_0$: Integration proceeds in two intervals 
$[z_e, z_0]$ and $[z_0, z_e]$ giving 
$$
I(z, z_e) = I_W(z)  + I_b(z_e), 
$$
where 
\begin{equation}
I_W(z) = i \frac{2}{3} \frac{|\omega|}{\sqrt{z_0}} (z - z_0)^{3/2}, 
\label{eq:ww1}
\end{equation}
\begin{equation}
I_b(z_e) = \frac{2}{3} \frac{|\omega|}{\sqrt{z_0}} (z_0 - z_e)^{3/2}
\label{eq:bb1}
\end{equation}

\item $z < \Delta z$, $z_e > z_0$: Integration is over a single interval $[z_e, z]$ with the result 
$$
I(z, z_e) = I_W(z)  + I_a(z_e), 
$$
where $I_W(z)$ is the same as in (\ref{eq:ww1}), and 
\begin{equation}
I_a(z_e) = - i \frac{2}{3} \frac{|\omega|}{\sqrt{z_0}} (z_e - z_0)^{3/2}. 
\label{eq:aa1}
\end{equation}
 
\item $z > \Delta z$ (narrow emission region), $z_e < z_0$:
$$
I(z, z_e) = I_N(z)  + I_b(z_e). 
$$
Here $I_b(z_e)$ is given in (\ref{eq:bb1}),  and $I_N(z)$ is the result of integration 
over two intervals $[z_0,\Delta z]$ and $[\Delta z,z]$: 
\begin{equation}
I_N(z) = i \frac{2}{3} \frac{|\omega|}{\sqrt{z_0}} 
\left[(\Delta z - z_0)^{3/2} + \frac{3}{2}(\Delta z - z_0)^{1/2} (z - \Delta z) \right]. 
\label{eq:nn1}
\end{equation}

\item $z > \Delta z$, $z_e > z_0$:
$$
I(z, z_e) = I_N(z)  + I_a(z_e)
$$
with $I_N(z)$ and  $I_a(z_e)$ defined in (\ref{eq:nn1})  and 
(\ref{eq:aa1}) respectively. 
\end{enumerate}
Using the expressions (\ref{eq:Omegaint3}) and (\ref{eq:Omegaint4}) 
and $p_\nu$ from (\ref{eq:psimple}), the solution (\ref{eq:Sint1}) can be written as 
\begin{equation}
  \label{eq:Sint3}
  S^{\rm int}(z) = -\frac{\sin2\theta_m}{\Delta z f_\nu(z)}
e^{-i [V_e z + I(z)]/s_\beta} 
\int_0^{z_{\rm max}} dz_e e^{-i [V_e z_e + I(z_e)]/s_\beta}, 
\end{equation}
where $z_{\rm max} = z$ for a wide region $z < \Delta z$, and 
$z_{\rm max} = \Delta z$ for a narrow region. The expressions for  
$f_\nu(z)$ and $I(z)$ should be taken according to the value of $z$. 

Since $I(z)$ is purely imaginary, we obtain  
\begin{equation}
  \label{eq:Sintfw}
  |S^{\rm int}(z)| = \frac{s_\beta \sin2\theta_m}{\Delta z f_\nu (z)} |G (z_{\rm max})| e^{-i I (z)/s_\beta}.  
\end{equation}
where 
\begin{equation}
  \label{eq:gee}
G(z_{\rm max}) \equiv \int_0^{z_{\rm max}} \frac{dz_e}{s_\beta} e^{-i [V_e z_e + I(z_e)]/s_\beta}. 
\end{equation}
The integral over $dz_e$  should be taken on the two intervals 
$[0, z_0]$ and $[z_0,  z_{\rm max}]$. In these intervals, $I(z_e)$ 
has the expressions  $I(z_e) = I_b(z_e)$ and $I(z_e) = I_a(z_e)$ respectively. 
This integral does not depend on whether the emission region is wide or narrow. 

In general, the integral over $z_e$ can not be computed analytically, but  
it can be estimated for  negligible $I(z_e)$ .
Averaging over the fast oscillations driven by $V_e$, we find for $z \gg z_0$   
\begin{equation}
\label{eq:Gz}
|G(z)| =  |G(\Delta z)| \approx  \frac{1}{V_e}.  
\end{equation}
Then the correlation from (\ref{eq:Sintfw}) becomes 
\begin{equation}
  \label{eq:Sintf}
  |S^{\rm int}(z)| = \frac{s_\beta\sin2\theta_m}{\Delta z  f_\nu(z) V_e} e^{-i I(z)/s_\beta}. 
\end{equation}

In what follows, we will consider 
the correlations in \eref{eq:Sintf} for wide and narrow regions separately 
and compare them with our numerical results.

\subsection{Narrow  emission region}
\label{sec:narrow2}

In the case of a narrow production region, $\Delta z < z_1$,  we have $f_\nu = 1$,
\begin{equation}
  \label{eq:Sintfn}
  |S^{\rm int}(z)| = \frac{s_\beta \sin2\theta_m}{\Delta z} |G(\Delta z)| e^{-i I_N(z)/s_\beta},  
\end{equation}
and the dependence on $z$ is in $I_N(z)$ only. Using the expression for $I_N(z)$ 
from (\ref{eq:nn1}) and $|G(\Delta z)| \approx 1/V_e$, we obtain  
\begin{equation}
  \label{eq:Sintzlarge}
|S^{\rm int}(z)| \approx \frac{s_\beta \sin 2\theta_m}{\Delta z V_e} 
\exp\left\{ \frac{2 |\omega| z_0}{3 s_\beta} \left[ \left(\frac{\Delta z}{z_0}-1\right)^{3/2} 
+ \frac{3 \Delta z}{2 z_0} \left(\frac{\Delta z}{z_0} - 1\right)^{1/2} \left(\frac{z}{\Delta z} 
- 1\right) \right] \right\}.
\end{equation}
The first term in the exponent can be written as $2\sqrt{2\mu \omega} \Delta z$. 
Therefore, for very narrow regions $\Delta z \ll 1/(2\sqrt{2 \mu \omega})$, it  can be neglected  
and the solution (\ref{eq:Sintzlarge}) becomes 
\begin{equation}
 |S^{\rm int}(z)| \approx \frac{s_\beta \sin 2\theta_m}{\Delta z V_e}  \exp
\left[s_\beta^{-1}\sqrt{|\omega| (2 \mu - |\omega|)}(z -  \Delta z) \right]. 
\label{eq:expon}
\end{equation}
The  correlations $|S^{\rm int}(z)|$  computed with (\ref{eq:Sintzlarge})  
for different values of $\Delta z$ are shown in \fref{fig:beam_lown}. 
There is a good agreement between the numerical and analytical results.  
The front factor of \eref{eq:expon} coincides with the expression for the averaged $\rho'_{12}$ 
in \eref{eq:rho12avsimple1}. Hence, the averaging effect can directly 
be seen in \eref{eq:expon} as a reduction of the initial value of $|S^{\rm int}|$ at $z=\Delta z$.

The result in (\ref{eq:Sintzlarge}) can be used in the interval $[\Delta z,  z_1]$. 
At $z > z_1$, the asymptotic behaviour starts which proceeds in the non-linear regime.  
It has a form of bi-polar oscillations with constant parameters which  
can be estimated as follows:   
For a thin neutrino sphere and  a constant value of $\mu$,
it was found in~\cite{Hannestad:2006nj}
that the probability in the  oscillation minima equals $P_{ee}^{\rm min} = \omega/ 2 \mu$.
At the maxima, $P_{ee} = 1$, so that the average equals $P_{ee} = 1 - \omega/ 4 \mu$. 
The period of bi-polar oscillations is 
\begin{equation}
l_{\rm bp} = 2 |\ln \theta|\frac{1}{\sqrt{\omega \mu}}. 
\label{eq:period}
\end{equation}  
These estimations are in very good agreement with the numerical results in \fref{fig:beam_lown}.

For a very narrow production region (\ref{eq:expon}),
the result can be obtained immediately from the corresponding evolution equation. 
Neglecting collective effects in the emission region, 
we can consider there the standard oscillations in matter with averaging. 
This gives the initial condition  for 
further exponential growth which starts at a certain distance $z_0'$.
Therefore, the emission profile is 
$$
p_\nu (z_e) = \delta (z - z_0'), 
$$
so effectively all the neutrino are emitted from the same point $z_0'$. 
In this case  the neutrino interaction term in the Hamiltonian (\ref{eq:Hsimple}) is  reduced to 
$\mu [\rho (z_0', t) - \bar{\rho}^* (z_0', t)]$. 
Plugging the matrices from (\ref{eq:defS}) into Eqs. (\ref{eq:simpleeom}) and (\ref{eq:Hsimple}) 
(and the corresponding equations for antineutrinos) 
and assuming  $\theta \approx 0$, we obtain  the following equations 
for $S$ and $\bar{S}$ to first order in $S$ and $\bar{S}$: 
\begin{equation}
  \label{eq:linS0}
  i s_\beta \frac{d}{dz}
  \begin{pmatrix}
    S \\ \bar{S}
  \end{pmatrix}
  = 
  \begin{pmatrix}
    -\omega + V_e - \mu & \mu \\ -\mu & \omega + V_e + \mu
  \end{pmatrix}
  \begin{pmatrix}
    S \\ \bar{S}
  \end{pmatrix}, 
\end{equation}
where $S = S(z_0', z)$. It can be obtained from  (\ref{eq:matrlin}) 
by replacing  $\tilde{\mu}$  by $\mu$.  
Following the same procedure as before, we arrive at the exponentially growing solution 
where the values of $Q$ and  $\bar{Q}$ are determined by the initial condition 
which takes into account averaging in the initial phase. 
$Q(z_0') =  2\langle \rho_{12}' (z_0) \rangle $ and for $|\langle\rho'_{12}(z_0)\rangle|$ 
we use the result given in (\ref{eq:rho12avsimple1}). 
Finally, 
\begin{equation}
  \label{eq:linsol}
  \left| \rho_{12}^{\nu,{\rm int}} \right| =  \frac{s_\beta \sin 2\theta_m }{2 \Delta z V_e} 
\exp \left[s_\beta^{-1} \sqrt{|\omega| (2 \mu - |\omega|)} (z - z_0')\right].
\end{equation}
For $z_0' = \Delta z$ this $\rho_{12}$  coincides with the result in (\ref{eq:expon}) up to a 
factor of two that comes from the choice of integration limits for \eref{eq:rho12avsimple1}. 

The very good agreement between the numerical solution and the linear solution 
demonstrates that the effect of an extended emission region on collective 
oscillations is
well described as an averaging of the initial state.

\subsection{Wide emission region}
\label{sec:largeDz}

In this case, $\Delta z > z_1$, and the instability develops completely  
inside the emission region. 
There are three regions of $z$ with different physics: 
$[0, z_1]$, where we can use results of linear approximation;  
$[z_1,  \Delta z]$, where the evolution becomes non-linear and 
non-trivial since more neutrinos are emitted and joint the system; 
$[ \Delta z, \infty]$ -- where the asymptotics appear. 

In the region   $z < z_1$, we have $f_\nu = z/\Delta z$ and according to (\ref{eq:Sintfw}),
\begin{equation}
  \label{eq:Sintfwwide}
  |S^{\rm int}(z)| = \frac{s_\beta \sin2\theta_m}{z} |G(z)|  e^{-i I_W(z)/s_\beta}. 
\end{equation}
Here the dependence on $z$ is more complicated: It appears in $I_W(z)$, $|G(z)|$ and in the pre-exponential factor. 
Using expression  (\ref{eq:ww1}) for $I_W(z)$ and \eref{eq:Gz} for $|G(z)|$, we find  
\begin{equation}
  \label{eq:Sintzsmall}
|S^{\rm int}(z)| \approx \frac{s_\beta \sin 2\theta_m}{z V_e} 
\exp\left[ \frac{2 |\omega| z_0}{3 s_\beta} \left(\frac{z}{z_0}-1\right)^{3/2} \right]. 
\end{equation}
Notice that here the exponential growth is  faster than in the case of  narrow regions: $\exp (A z^{3/2})$. 
The reason is that more neutrinos join the system and $\tilde{\mu}$ 
increases in the range of $z$ where the instability develops.  
In \fref{fig:beam_highn}, the analytic results from \eref{eq:Sintzsmall} are  shown as bold lines. 
They agree  with the numerical solutions. 
In particular, \eref{eq:Sintzsmall} captures well the changing slope with $z$ for a 
given $\Delta z$ as well as the different slopes for different $\Delta z$. 
Clearly, it is important to use $\tilde{\mu}$ rather than $\mu$.

Some differences between the numerical and analytical results 
seen in  \fref{fig:beam_highn} are related  to  assumptions and approximations 
involved in deriving \eref{eq:Sintzsmall}. 
In particular, 
\begin{enumerate}
\item  In the  ansatz (\ref{eq:Ssol}), we neglected the $z_e$ dependence of $Q$. 
It  appears via  dependence of mixing on $\tilde{\mu}(z_e)$.  
\item The $z_e$-dependent part, $I(z_e)$, has been neglected  in the integral over $z_e$ 
when making the approximation in (\ref{eq:Gz}).
\end{enumerate}
Both approximations have an accuracy of the order  $\mu/V_e \sim  10\%$.\\

Let us consider the evolution in the interval $[z_1, \Delta z]$ in the non-linear 
regime but before the system reaches the asymptotics. 
Here we cannot use the results of the linear approximation. 
Now the collective oscillations  are  not as efficient as in the case $z_1 > \Delta z$. 
Indeed, according to \fref{fig:beam_highn}, the depth of the oscillations 
becomes smaller and the minima become shallower in comparison with the narrow region case. 

Generalising the result for the probability at the minimum 
found in~\cite{Hannestad:2006nj}  (see \sref{sec:narrow2}), 
we substitute $\mu$ by $\tilde{\mu}$ in the first oscillation minimum $z_{\rm min}$:  
\begin{equation}
  \label{eq:firstdip}
  \rho_{ee}^{\rm int}(z_{\rm min}) \approx \frac{|\omega|}{2\tilde{\mu}(z_{\rm min})}.
\end{equation}
This expression is in very good agreement with results of numerical computations.  
The averaged value of $P_{ee}$ decreases, converging to the
asymptotic behaviour at $z  > \Delta z$,  where the oscillations proceed around the average value  
\begin{equation}
  \label{eq:rhoasym}
  \langle \rho_{ee,\mathrm{asymptotic}}^{\rm int} \rangle
\approx \frac{1}{2}\left(1 + \frac{|\omega|}{2\mu} \right)
\end{equation}
given by the total $\mu$ being independent of the width of the production region.

\begin{figure}[tbp]
  \centering
  \includegraphics[width=0.8\textwidth]{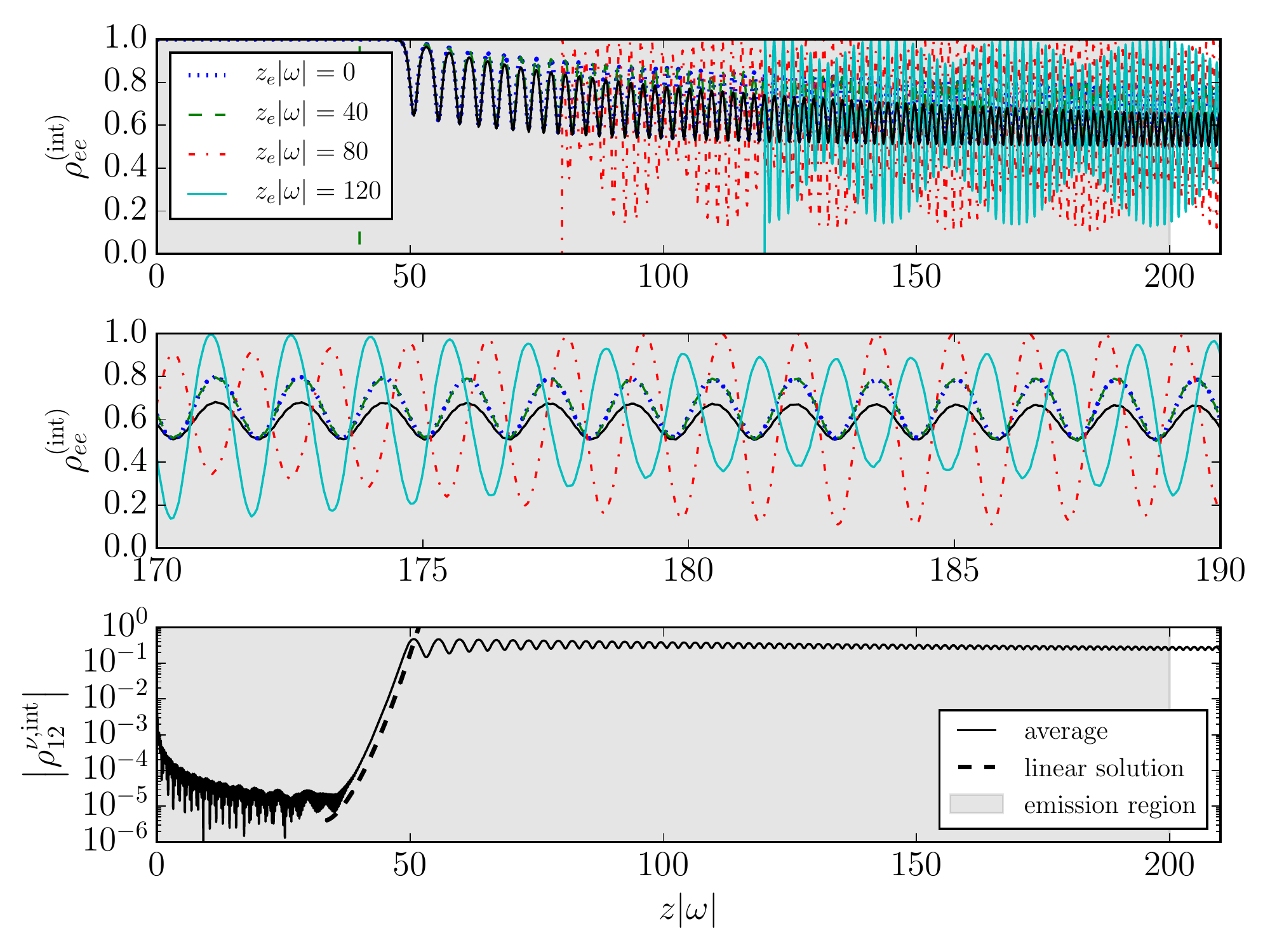}%
  \caption{Collective flavour conversion within a very large emission region $\Delta z |\omega| = 200$. 
{\it Upper panel:} The  $\nu_e$ survival probabilities  averaged  over neutrino production points 
(solid line) as well as  for neutrinos produced in different points as function of distance. 
{\it Middle panel:} A zoom  of the probabilities in the range  $z |\omega| = 170$ to $190$. 
{\it Bottom panel:}  The off-diagonal term of  the density matrix and the linear solution 
according to \eref{eq:Sintzsmall}.  
The shading indicates the emission region.}
\label{fig:beam_vhighn}
\end{figure}

In order to understand  this behaviour, 
we present results for a very large emission region, $z|\omega| = 200$ in \fref{fig:beam_vhighn}. 
In this case, $z_1 \approx 50$ and the onset coordinate $z_0 = 33$.  In the lower panel 
one sees  again a good agreement between the analytic 
\eref{eq:Sintzsmall} (the dashed line) and the numerical results (the solid line). 
Furthermore, the value of $\rho_{ee}$ at the minimum is in very good agreement  with  
$\rho_{ee} (z_{\rm min }) = 0.66$  predicted by \eref{eq:firstdip}.

In the upper panel of \eref{eq:firstdip} 
we show the oscillation probabilities  for  individual 
neutrinos emitted at certain  points: $z_e |\omega| = 0, 40, 80$ and $120$ 
in addition to the probability averaged over all the neutrinos emitted before a given point $z$.

Neutrinos emitted at 0 and 40, that is in the linear regime $z_e < z_1$, 
have almost identical flavour evolutions, and the phases of bi-polar oscillations agree 
with that of the averaged one. 
This is essentially the consequence
of factorisation of dependencies on $z$ and $z_e$, in which
all the modes (produced at different $z_e$)
have the same dependence on $z$ while differences in phases
related to $z_e$ are encoded in the suppression factor given by the integral
$|G(z)|$. So, the phases of their bi-polar
oscillations are in a sense synchronised.

Neutrinos produced after $z_1$ ($z_e > z_1$) start the bi-polar
oscillations immediately from $z_e$. The initial phases of these
oscillations are determined  by the fact
that  $\nu_e$ and $\bar{\nu}_e$ are emitted.
These phases are not related to the phase of the collective bi-polar
oscillation of the rest of the system (formed by neutrinos emitted earlier).
Furthermore, the frequencies of bi-polar oscillations increase with $\mu'$ as
$$
\omega_{\rm bp} \sim \sqrt{|\omega|\mu'},
$$
where $\mu'$ is some effective value of $\mu$ found by averaging over the other neutrinos.

In \fref{fig:beam_vhighn} one can see the difference of the averaged bi-polar frequencies of modes produced in the linear regime ($z_e < z_1$) and modes produced in the non-linear regime ($z_e > z_1$). Frequencies within these two groups are the same.

All neutrinos produced in the linear regime have almost identical flavour evolutions, and due to the commutator structure of \eref{eq:simpleeom}, a neutrino is not affected by a background neutrino if the flavour states are identical. Hence, neutrinos produced in the linear regime effectively see a slightly lower background of neutrinos than the neutrinos produced in the non-linear regime, and this results in the lower frequency.

As a consequence,  neutrinos with $z_e = 80$ and $120$ are in phase with the whole ensemble at some values of $z$, and at these points, their oscillations are enhanced. 
At other points, they are out of phase, and the oscillations are smaller in amplitude. 
This is seen in the central (zoomed) panel of \fref{fig:beam_vhighn}.  
The total ensemble  also reflects this effect of late neutrinos in a certain way.  Indeed,
in absence of collective effects, the  oscillations of  neutrinos produced after $z_1$  would
average, so that their contribution to the depth of oscillations
would be negligible being suppressed by a factor $(\frac{3}{4} \Delta z \omega_{\rm bp})^{-1}$.
In this case the depth of the bi-polar oscillations at the end of the production region equals
$$
d(\Delta z) \approx \frac{1}{4} d(z_1), 
$$
where $d(z_1)$ is the depth at the beginning of bi-polar
oscillations. The factor $1/4$ originates  from the normalisation of the density matrix
since $z_1 \approx \Delta z/4$, and before $z_1$ only 1/4 of all neutrinos are emitted.
In \fref{fig:beam_vhighn}, we observe that the final amplitude is larger than $\frac{1}{4} d(z_1)$, and this is a result of the enhancement described above for neutrinos emitted at $z_e > z_1$.

\section{Towards the realistic  case}
\label{sec:getreal}

There are two directions to approach the realistic situation in SN: 
(i) Use the same model (profiles, emission) as in \sref{sec:linear} but change the parameters.  
(ii) Modify the model. 

\subsection{Variation of  parameters}

The analytic results obtained in the previous section, 
which describe well the results of the numerical computations,  can be used to extrapolate 
results to a more realistic situation.
Although we do not expect to describe the physics arising in the neutrino 
production region of a supernova since we have neglected,
 e.g.,  multi-angle effects~\cite{EstebanPretel:2007ec,EstebanPretel:2008ni}, as well as the different densities of $\nu_e$ and $\bar{\nu}_e$,
this still gives an idea about how the numbers relate to realistic densities and length scales.

We approximate the profile of neutrino emission of \fref{fig:prob}
by a box-like dependence which matches our model of \sref{sec:linear}. 
Then $\Delta z = 5$ km and  $\Delta z \omega = 3.14$. 
We use 
\begin{equation}
V_e = 10^{8}|\omega|, \quad \mu = 0.1 V_e = 10^{7} |\omega|, 
\end{equation}
which correspond  to the middle of the production region in \fref{fig:prob} 
(compare with \eref{eq:parameters}). 
With these values of the parameters, the coordinate of the onset 
of the instability (\ref{eq:onset}) equals  
$$
z_0 = 5 \cdot 10^{-8} \Delta z
$$
or $z_0 = 2.5 \cdot 10^{-2}$ cm, and $\omega z_0 = 1.6 \cdot 10^{-6}$, {\it i.e.} much smaller than 
the vacuum oscillation length. 
So, $z_0 \ll \Delta z$ and the exponential growth starts practically 
at the very beginning of the emission region. 
The mixing angle in matter is $\sin 2 \theta_m = 10^{-9}$, and the  
factor in front of the exponent (e.g. in \eref{eq:Sintzlarge}), which 
includes smallness of the initial mixing and the suppression of correlation due to averaging, equals 
$$
D \equiv \frac{s_\beta \sin 2 \theta_m}{\Delta z V_e} = 2.3 \cdot 10^{-18}. 
$$

The coordinate of complete development of collective oscillations is determined by the condition 
$|S^{\rm int}(z_1)| = 1$,  and using (\ref{eq:Sintzsmall}) for the wide emission region, we obtain 
\begin{equation}
  \label{eq:z1det}
z_1 \approx z_0 \left[ \frac{3 s_\beta}{2 |\omega| z_0} 
\left( \ln D^{-1} - \ln \frac{\Delta z}{z_1^*} \right) \right]^{2/3} = 
\left[\frac{3 s_\beta}{2} 
\left( \ln D^{-1} - \ln \frac{\Delta z}{z_1^*} \right)
\right]^{2/3} 
\left(\frac{\Delta z} {2 |\omega| \mu}\right)^{1/3}.  
\end{equation}
Here, $z_1^* = z_1 (z = \Delta z)$ is  determined by \eref{eq:z1det} with zero 
last term in the brackets: 
\begin{equation}
  \label{eq:z1detst}
z_1^* = 
\left(\frac{3 s_\beta}{2} \ln D^{-1} \right)^{2/3}
\left(\frac{\Delta z} {2 |\omega| \mu}\right)^{1/3}.
\end{equation}
Numerically, we find 
$$
z_1 = 10^4 {\rm cm } \left(\frac{\Delta z}{ 5~{\rm km}}\right)^{1/3} 
\left(\frac{\mu}{0.1V_e}\right)^{- 1/3} \left(\frac{|\omega|}{10^{-8}V_e}\right)^{- 1/3} 
$$
and $z_1/\Delta z = 0.022$. 
So, indeed,  $z_1 \ll \Delta z$  and the assumption about  wide emission case is justified. 

To quantify the effect of integration  over the production region, we compare the result 
(\ref{eq:z1det}) with the one with surface emission (\ref{eq:linsol}) 
where we also eliminated suppression due to averaging:
\begin{equation}
  \label{eq:linsol3}
  \left| S^{\rm int} \right| =  s_\beta \sin 2\theta_m
\exp\left[\frac{|\omega| z_0}{s_\beta} 
\left(\frac{2\mu}{|\omega|} - 1\right)^{1/2} \left(\frac{z}{z_0}-1\right) \right]  .
\end{equation}
This equation and the condition  $\left| S^{\rm int}(z_1) \right| = 1$ give  
\begin{equation}
  \label{eq:z1det2}
z_1^\delta \approx  \frac{s_\beta}{\sqrt{2|\omega| \mu}} \ln(s_\beta \sin 2\theta_m)^{-1}. 
\end{equation}
Numerically, we obtain $z_1 =  5 \cdot 10^{2}$ cm. 
This is smaller than $z_1$ found for the wide emission region by a factor 20.   

\subsection{Modifying the model}

In the simple model, we can understand both the onset 
of flavour conversion and the asymptotic behaviour. However, the treatment of the production 
is rather crude. 
One of the interesting features that will be worthwhile to explore with 
an improved treatment is the convergence towards $\rho_{ee,\textrm{asymptotic}}$.
Such convergence can e.g. be relevant for 'fast flavour conversion' inside the 
neutrino production region~\cite{Dasgupta:2016dbv, Capozzi:2018clo}.

In the present study, we have made a number of simplifications  
that can affect the validity of our results for real SNe, 
and in the following, we will discuss the most important of them.

\begin{itemize}

\item 
We considered  production and averaging of electron (anti)neutrinos only. 
For muon and tau neutrinos, a similar averaging can be performed, 
and the final result is the sum of the individual density matrices.
The width of the muon- and tau neutrino emission profiles are similar to the width of the electron 
neutrino emission profile, so the results will be unchanged within an order of magnitude.

\item
We assumed that the neutrino density is the smallest at $z=0$ 
and it increases  as the neutrinos are produced throughout the emission region. 
In some sense this is the opposite to the situation in supernovae where the neutrino density 
is largest at the smallest radii. Essentially one should take into account neutrinos 
which are both emitted and absorbed in the production. They influence the evolution of neutrinos which will escape.  
Notice, however, that at small radii, all three neutrino species are in equilibrium, 
such that $\rho_{ee} \approx \rho_{\mu\mu} \approx \rho_{\tau\tau}$ (assuming a small lepton number which is a good approximation during the cooling phase). 
As a consequence, the traceless part of $\rho$ which affect neutrino oscillations 
is comparatively small at small radii and starts to grow only when $\rho_{\mu\mu}$ and 
$ \rho_{\tau\tau}$ become different from $\rho_{ee}$. Effectively this corresponds to a 
growth of $\mu$. 

\item 
We consider only one energy for the neutrinos. In the cases where  the 
neutrino background is ignored, this assumption does not affect results. In fact,  we expect that  
modes with different energies  behave similarly even in the presence of collective oscillations, 
so our main results are expected to be reproduced for continuous spectra of neutrinos.

\item
We consider equal fluxes of $\nu_e$ and $\bar{\nu}_e$ which removes the synchronised oscillations which would otherwise dominate at high neutrino densities~\cite{Pastor:2001iu}. However, the suppresion we found is still expected to be present, and only the large amplitude flavour conversion is expected to be delayed until lower neutrino densities.

\item Constant $V_e$ and $\mu$ (outside the production region) have been considered. 

\item The model has a simple geometry with a single emission angle.
Inclusion of multi-angle matter effects~(e.g. \cite{EstebanPretel:2007ec,EstebanPretel:2008ni})
can suppress or even remove the instability, 
and in this case additional suppression due to averaging 
has no significant impact. 
However, in some models, collective instabilities emerge at larger radii, and the averaging  effect 
we found will still be present and important.

\item 
In our simple model, we  introduce 
a distribution of neutrino production points. 
This  captures most of the important  physics.  
However, a complete and consistent description  of neutrino production and oscillation
would require solution of  the full quantum kinetic equations (QKE), especially to get
the correct asymptotic solutions which may be affected by neutrinos that are
emitted and then reabsorbed.   
A very recent study takes a significant step towards such computations \cite{Richers:2019grc}.

\end{itemize}
Thus, our results can be generalised in several different directions, 
but none of them is expected to change the overall conclusions.

\section{Conclusions}
\label{sec:conclusions}

\begin{itemize}

\item We explored the effect of a finite width of the neutrino production region
in SNe on collective flavour transformations.
We find that the effective width of this region $r_{\rm eff} \sim 5$ km
is much larger than the oscillation length given by $V_e$:
$D \propto 1/ r_{\rm eff} V_e \sim 10^{-10} - 10^{-8}$.
Averaging over the production region leads
to the additional suppression of conversion effects in the initial phase
of the order of $D$. 
Thus, the usually used assumption of
emission from a neutrino sphere with infinitesimal width
justified by strongly suppressed
mixing in matter is flawed.

\item The most transparent and adequate consideration of
dynamics of conversion in the presence of averaging is given
in terms of the correlation $\rho_{12}$ -- the off-diagonal element
of the density matrix in basis of the eigenstates
of the Hamiltonian. In the phase of developments of instability
$|\rho_{12}| \approx |\rho_{ex}|$.
It is $\rho_{12}$ that gets the additional suppression factor $D$
due to averaging and it is  $\rho_{12}$
that determines the onset of collective oscillations.
We show that $\rho_{12}$ with the additional suppression factor should be used
as the initial condition for further collective transformations.
In the phase of evolution when averaging and development of the instability occur,
one can use a linear approximation for $\rho_{12}$.
We derived the evolution equations for the averaged $\rho_{12}$ in this approximation.
We found the analytic solutions of these equations in the simplified model
of two intersecting fluxes and uniform emission of neutrinos in the region
$\Delta z$. The analytic solutions are in very good agreement with the results of
numerical computations.

\item Averaging over the production region
does not eliminate the development of instabilities, but it
leads to a delay of strong collective effects (to increase of $z_1$).
The delay depends on features of the development of the instability.
In the case of a narrow production region and exponential growth of
correlation $\rho_{12}$, the delay is given by the logarithm of
the additional suppression factor, $\ln D^{-1}$.
For a wide production region (when the complete development of
the instability occurs inside the production region),
averaging modifies the exponential growth. In our example, it becomes
$\exp A z^{3/2}$. In this case the averaging also modifies bi-polar oscillations 
in the non-linear regime and asymptotics.

\item Our qualitative results concerning the averaging are rather generic
in spite of the fact that they were obtained in a framework of simplified
models. In particular,  if strong transformations occur
outside the production region, the additional suppression
factor due to averaging should be included in the initial condition
for further evolution. That leads to the delay of strong transformations
which depends on the specific form of the collective effect.
If strong transformation develops inside the production region,
averaging leads to a modification of the instabilities' growth
and asymptotics in addition to the delay.
\end{itemize}

In conclusion,  there are three formal results in the paper.
The first one is that \emph{averaging due to different emission points 
can decrease the off-diagonal part of the density matrix for the propagating 
neutrino states by many orders of magnitude.} The reason is that the phases 
of neutrinos emitted at different points in the supernova are independent. 
The width of the neutrino sphere is $\sim 8-10$ orders of magnitude 
larger than the neutrino oscillation length, and this factor determines the suppression of  
the off-diagonal part of the density matrix.

The second result  is that \emph{linear stability analysis should be done 
in the basis of propagating states.} This observations is tightly connected 
to the first result concerning averaging. 
We find that the growth rates determined in the flavour basis and in the basis of propagating 
states are the same. The main difference appear in the identification of  
the onset of growth which is problematic in the flavour basis.

The third result is that \emph{averaging gives the
correct initial condition for constructing a linear solution.}
Using this result, it will be possible in the future to take into account 
the effect of an extended neutrino sphere on collective neutrino conversion 
that occurs well outside the neutrino sphere. For conversion taking place 
inside the neutrino sphere, our result will also help to find  appropriate 
initial conditions, but a full solution of the QKE is needed to determine
the asymptotic behaviour.

\section*{Acknowledgements}
R.S.L.H. would like to thank Steen Hannestad and Irene Tamborra for helpful discussions. 
R.S.L.H. was partly funded by the Alexander von Humboldt 
Foundation. The work of A.S. is supported by Max-Planck senior fellow grant M.FW.A.KERN0001.

\appendix 

\section{Numerical implementation}
\label{sec:numerical}

In order to numerically follow the evolution of $\rho$, \eref{eq:eom} for $\psi$with $r$ replaced by $t$, and $\rho$ is calculated as $\rho = \psi^\dagger \psi$ which is entirely equivalent to solving \eref{eq:simpleeom} with the condition $\rho = \rho^\dagger$. The evolution is followed as a function of $z$ by using the relation $dz/dt = \sin\beta$. 

The equation is solved with a complex-valued variable-coefficient ordinay differential equation solver with fixed-leading-coefficient implementation using an implicit Adams method\footnote{\url{http://www.netlib.org/ode/zvode.f}}. Good convergence was found for all calculations when using an absolute tolerence of $10^{-14}$ and a relative tolerance of $10^{-10}$. 

The finite emission region was implemented by using $N_e$ bins for the emission point $z_e$. Good convergence was found for $\Delta z V_e / 2\pi > N_e$ which corresponds to resolving the fastest oscillations. 
For the calculation in \fref{fig:beam_lown}, 141 bins were used. In \fref{fig:beam_highn}, $N_e=401$, and in \fref{fig:beam_vhighn}, $N_e=2000$. In each case the convergence was confirmed by running a test with $N_e$ larger than that as well as several tests with lower resolution. At points where $z<z_e$, $\psi$ is put to zero in order to only start the evolution of the neutrino flavour after the neutrino is emitted.

\section{Transformation from $\rho$ to $\rho^\nu$}
\label{sec:transform}

The density matrix in the basis of the eigenstates of the Hamiltonian $\rho^\nu$ 
and  in the flavour basis  $\rho_f$ are related as 
\begin{equation}
  \label{eq:transform12}
  \rho^\nu =  U (\theta_\nu)^{\dagger}  \rho_f U(\theta_\nu),      
\end{equation}
where $U(\theta_\nu)$ is the mixing matrix in matter (\ref{eq:mixing}) 
where the neutrino background is accounted for. 
The the off-diagonal element of $\rho^\nu$ can be written explicitly as
\begin{equation}
  \label{eq:transformflavourfull}
  \rho_{12}^\nu = \frac{1}{2} \sin 2\theta_\nu (\rho_{ee} - \rho_{xx}) - 
\sin^2 \theta_\nu \rho_{xe} + \cos^2\theta_\nu \rho_{ex}.
\end{equation}
We assume that the matter potential dominates ($V_e \gg \mu, \omega)$, 
which gives $\theta_\nu \approx \pi/2$ for normal mass ordering (NO) 
and $\theta_\nu \approx 0$ for inverted mass ordering (IO). 
In the  NO case, the last term in \eref{eq:transformflavourfull} is negligible. 
For $|\rho_{xe}| \sim 1$, the first term is negligible due to the smallness of $\sin 2\theta_\nu$, 
In opposite case  $|\rho_{ex}| \ll 1$, we have $\rho_{ee} \approx 1$, $\rho_{xx} \approx 0$ 
and the first term is approximately $\frac{1}{2} \sin 2\theta_\nu$. Combining these two limits, we find
\begin{equation}
  \label{eq:transformflavourNO}
  \rho_{12}^\nu \approx \frac{1}{2} \sin 2\theta_\nu - \rho_{ex}^*
\end{equation}
for NO, and
\begin{equation}
  \label{eq:transformflavourIO}
  \rho_{12}^\nu \approx \frac{1}{2} \sin 2\theta_\nu + \rho_{ex}
\end{equation}
for IO. 
The value of $\sin 2\theta_\nu $ can be determined 
considering the off-diagonal elements of the Hamiltonian (\ref{eq:Hsimple})
through the relation
\begin{equation}
  \label{eq:sin2thetanu}
  \omega_\nu \sin 2\theta_\nu = \omega \sin 2\theta + 
2 \mu \int_0^{z} (\rho_{ex}(z_e',z) - \bar\rho_{ex}^*(z_e',z)) p_{\nu}(z_e')dz_e'.
\end{equation}
In turn, to determine approximate values for $\rho_{ex}$ and $\bar\rho_{ex}$, 
we put $\mu = 0$ 
and assume that oscillations are averaged so that $\rho_{12}^\nu=\bar{\rho}_{12}^\nu=0$. 
This gives 
\begin{equation}
  \rho_{ex}  \approx \frac{\omega \sin 2\theta}{2(V_e-\omega \cos2\theta)}, ~~~~
\bar{\rho}_{ex}  \approx -\frac{\omega \sin 2\theta}{2(V_e+\omega \cos2\theta)},
\end{equation}
where the sign difference arises because the matter term in the Hamiltonian has 
opposite signs for neutrinos and antineutrinos, and the approximation 
$\omega_\nu \approx  V_e \mp \omega \cos2\theta$ was used.
Since $\rho_{ex}$ and $\bar{\rho}_{ex}$ are independent of $z_e'$, 
the integral in \eref{eq:sin2thetanu} can be absorbed in the definition of $\mu$ as in 
(\ref{eq:mutildedef})
\begin{equation}
  \tag{\ref{eq:mutildedef}}
  \tilde{\mu}(z) = \mu f_\nu(z).
\end{equation}
Finally, with a Taylor expansion of $\rho_{ex}$ and $\bar\rho_{ex}$ in $\omega \cos 2\theta/V_e$, 
we arrive at \eref{eq:rhonutilde}.

\section{Linear equations for $\rho^\nu$ in the eigenstate basis}
\label{sec:eigenstate}

Here we  find the evolution equation for the off-diagonal elements of the density 
matrix in the eigenstate basis. 
Let us take that at the emission  $\rho = \bar{\rho}$. 
As far as  $\rho_{ex}, \bar{\rho}_{ex} \ll 1$, 
the Hamiltonian in the flavour basis (\ref{eq:Hsimple}) 
can be written in the lowest order in  $\rho_{ex}, \bar{\rho}_{ex}$ as 
\begin{equation}
  \label{eq:Heq}
  H = \frac{\omega}{2}
  \begin{pmatrix}
    -c_{2\theta} & s_{2\theta} \\ s_{2\theta} & c_{2\theta}
  \end{pmatrix}
  + 
  \begin{pmatrix}
    V_e & 0 \\ 0 & 0
  \end{pmatrix}
  + \mu \int dz_e' p_{\nu}(z_e')
  \begin{pmatrix}
    0 & \rho_{ex}-\bar{\rho}_{ex}^* \\  \rho_{xe}-\bar{\rho}_{xe}^* & 0
  \end{pmatrix} .
\end{equation}
Let us determine the eigenstates of this Hamiltonian which depends on the density matrices. 
The latter produces a complication in comparison to the case without $\nu \nu$ interactions. 
The problem can be treated in the following way. 
Let us introduce the 'fixed  point' density matrix $\rho^0$ for neutrinos and $\bar{\rho}^0$
for antineutrinos which are constant in time and are determined by the self consistency 
conditions $[H^0,\rho^0]=0$ and $[\bar{H}^0,\bar{\rho}^0]=0$, where $H^0$ and $\bar{H}^0$ are
the Hamiltonians with $\rho$ and $\bar\rho$ substituted by the fixed point matrices.
Diagonalisation of $H^0$ determines 
the eigenstate basis which is related to the flavour basis by the angle $\theta_\nu$ and the eigenvalues. 
The difference of the eigenvalues (frequency of oscillations) equals 
\begin{equation}
  \label{eq:omeganu}
  \omega_\nu = \sqrt{(V_e-\omega c_{2\theta} )^2 + \left|\omega s_{2\theta} + 
2 \mu \int (\rho_{ex}^0 - \bar{\rho}_{ex}^{0*}) p_{\nu}(z_e')dz_e'\right|^2}.
\end{equation}
The off-diagonal parts of $\rho^0$ and $\bar{\rho}^0$ in the flavour basis, $\rho^0_{ex}$ and 
$\bar{\rho}^0_{ex}$, are expected to be small, but non-zero. For the neutrino, $\rho^0_{ex}$ 
can be determined by putting $\rho^\nu_{12}=0$ in (\ref{eq:transformflavourNO}) or
(\ref{eq:transformflavourIO}), and $\bar{\rho}^0_{ex}$ can be determined in a similar way.
In terms of $\theta_\nu$ and  $\omega_\nu$, the Hamiltonian $H^0$ is represented as 
\begin{equation}
  \label{eq:defnu}
  H^0 = \frac{\omega_\nu}{2} 
  \begin{pmatrix}
    -c_{2\theta_\nu} & s_{2\theta_\nu} \\ s_{2\theta_\nu} & c_{2\theta_\nu}
  \end{pmatrix}.
\end{equation}
Then the full Hamiltonian (\ref{eq:Heq}) can be  rewritten as 
\begin{equation}
  \label{eq:Hgen}
  H = \frac{\omega_\nu}{2} \hspace{-1bp}
  \begin{pmatrix}
    -c_{2\theta_\nu} & s_{2\theta_\nu} \\ s_{2\theta_\nu} & c_{2\theta_\nu}
  \end{pmatrix}
  + \hspace{-0.5bp} \mu \hspace{-5bp} \int dz_e' p_{\nu}(z_e')
  \begin{pmatrix}
    0 & \rho_{ex}\hspace{-1bp}-\hspace{-1bp}\bar{\rho}_{ex}^* \hspace{-1bp}-\hspace{-1bp} (\rho_{ex}^0\hspace{-1bp}-\hspace{-1bp}\bar{\rho}_{ex}^{0*})\\  
\rho_{xe}\hspace{-1bp}-\hspace{-1bp}\bar{\rho}_{xe}^* \hspace{-1bp}-\hspace{-1bp}(\rho_{xe}^0\hspace{-1bp}-\hspace{-1bp}\bar{\rho}_{xe}^{0*}) & 0
  \end{pmatrix}.
\end{equation}
Similarly, we find the Hamiltonian for antineutrinos 
\begin{equation}
  \label{eq:Hbar}
  \bar{H} = \frac{\omega_{\bar{\nu}}}{2}
  \begin{pmatrix}
    -c_{2\theta_{\bar{\nu}}} & s_{2\theta_{\bar{\nu}}} \\ 
s_{2\theta_{\bar{\nu}}} & c_{2\theta_{\bar{\nu}}}
  \end{pmatrix}
  - 
  \begin{pmatrix}
    V_e & 0 \\ 0 & 0
  \end{pmatrix}
  + \mu \int (\bar{\rho} - \rho^*  - (\bar{\rho}^0 - \rho^{0*}) ) p_{\nu}(z_e')dz_e',
\end{equation}
where 
\begin{equation}
  \label{eq:omeganubar}
  \omega_{\bar{\nu}} = \sqrt{(V_e+\omega c_{2\theta} )^2 + \left| \omega s_{2\theta} + 
2 \mu \int (\bar{\rho}_{ex}^0 - \rho_{ex}^{0*})p_{\nu}(z_e')dz_e' \right|^2},
\end{equation}
and  $\theta_{\bar{\nu}} \neq \theta_\nu$.

Let us find the Hamiltonians in the eigenstate basis. 
Transformation of  $H$ in \eref{eq:Hgen} with the  matrix $U_{\theta_\nu}$ gives
\begin{equation}
  \label{eq:Hnu}
  H^\nu = \frac{\omega_\nu}{2}
  \begin{pmatrix}
    -1 & 0 \\ 0 & 1
  \end{pmatrix}
  + \mu \int dz_e' p_{\nu}(z_e') \left[\rho^\nu - V^\dagger \bar{\rho}^{\bar{\nu}*} V - 
(\rho^{\nu 0} - V^\dagger \bar{\rho}^{\bar{\nu}0*} V )\right] ,
\end{equation}
where $V = U_{\theta_{\bar{\nu}}}^\dagger U_{\theta_{\nu}}$. 
Taking the limit of $V_e \gg \omega$, we find to lowest order in $\omega/V_e$ for NO
\begin{equation}
  \label{eq:cossin}
  c_{\theta_\nu} \approx \frac{1}{2}\left(1+ 2 \frac{\mu}{V_e}\right) \epsilon, \quad s_{\theta_\nu} \approx 1, 
\quad c_{\theta_{\bar{\nu}}}\approx 1, \quad s_{\theta_{\bar{\nu}}} \approx \frac{1}{2}\left(1-2 \frac{\mu}{V_e}\right) \epsilon ,
\end{equation}
where $\epsilon \equiv  |s_{2\theta} \omega/V_e|$.
The corresponding matrix is then 
\begin{equation}
  \label{eq:Veq}
  V \approx 
  \begin{pmatrix}
    \epsilon &  1 \\ - 1 & \epsilon
  \end{pmatrix} .
\end{equation}
For IO $V \rightarrow -V$, but the overall sign will not be important in the following.
In general,  we find
\begin{equation}
  \label{eq:Vaction}
  V^\dagger \rho V = 
  \begin{pmatrix}
    \rho_{22} - \epsilon (\rho_{12}+\rho_{21}) & -\rho_{21} + \epsilon (\rho_{11}-\rho_{22})\\
    - \rho_{12} + \epsilon (\rho_{11}-\rho_{22})& \rho_{11} + \epsilon (\rho_{12}+\rho_{21})
  \end{pmatrix} .
\end{equation}
In order to linearise the evolution equations for $\rho$, we use the definitions in \eref{eq:defS}.
Assuming $\bar{s}=s$, the term $\rho^{\nu 0} - V^\dagger \bar{\rho}^{\bar{\nu}0*} V$ 
cancels the off-diagonal $\epsilon$ terms that arise from the term $V^\dagger \bar{\rho}^{\bar{\nu}} V$ 
according to \eref{eq:Vaction}, and the linearised equations are given by
\begin{equation}
  \label{eq:Slin}
  \begin{aligned}
    i s_\beta\frac{dS}{dz} &= \omega_\nu S + \mu \int (s' S - S' s - \bar{s}' S + \bar{S}' s)p_{\nu}(z_e')dz_e',\\
    i s_\beta\frac{d\bar{S}}{dz} &= \omega_{\bar{\nu}} \bar{S} + \mu \int (-\bar{s}' \bar{S} + \bar{S}' \bar{s} + 
s' \bar{S} - S' \bar{s})p_{\nu}(z_e')dz_e',
  \end{aligned}
\end{equation}
which leads to \eref{eq:linSgen} when $s$, $\bar{s}$, $\omega_\nu$ and $\omega_{\bar{\nu}}$ are inserted.

In the limit $V_e \gg \omega$, we get $\omega_\nu = V_e - c_{2\theta}\omega $ and $\omega_{\bar{\nu}} 
= V_e + c_{2\theta}\omega $. 
Often it is argued that a large $V_e$ suppresses the mixing angle in collective oscillations, 
however, here it is clear that this is only partly true given the factor $c_{2\theta}$ in front 
of $\omega$.
Instead, the approximation has to be justified based on the vacuum mixing angle. 
In the three-neutrino case this has even more noticeable consequences~\cite{Doring:2019axc}.

\end{document}